\pdfoutput=1

\documentclass[11pt]{article}

\usepackage[final]{acl}

\usepackage{times}
\usepackage{latexsym}

\usepackage[T1]{fontenc}

\usepackage[utf8]{inputenc}

\usepackage{microtype}

\usepackage{inconsolata}

\usepackage{graphicx}

\usepackage{algorithm}
\usepackage{algorithmic}
\usepackage{xcolor}
\usepackage{array}
\usepackage{makecell}
\usepackage{afterpage}
\usepackage[table]{xcolor}

\usepackage{comment}
\usepackage{booktabs}
\usepackage{multirow}
\usepackage{amsmath}
\usepackage{caption}

\usepackage{hyperref}
\definecolor{darkblue}{rgb}{0,0,0.6}
\hypersetup{
    colorlinks=true,
    linkcolor=darkblue,
    citecolor=darkblue,
    urlcolor=darkblue,
    pdfborder={0 0 0}
}
\usepackage[utf8]{inputenc}
\usepackage{amsfonts}
\usepackage{dsfont} 
\newcommand{\indic}[1]{\mathds{1}\!\left[#1\right]}
\usepackage{enumitem}
\usepackage[normalem]{ulem} 

\newif\ifcolordiff
\colordifffalse   

\ifcolordiff
  \newcommand{\added}[1]{\textcolor{blue}{#1}}
  
  \newcommand{\deleted}[1]{\textcolor{red}{\sout{#1}}}
\else
  \newcommand{\added}[1]{#1}
  
  \newcommand{\deleted}[1]{}
\fi

\setlist[itemize]{topsep=2pt,partopsep=0pt,itemsep=2pt,parsep=0pt}
\setlist[enumerate]{topsep=4pt,partopsep=0pt,itemsep=4pt,parsep=0pt}

\begin{document}

%
%

\title{Private Seeds, Public LLMs: Realistic and Privacy-Preserving \\Synthetic Data Generation}


\author{
  \textbf{Qian Ma},
  \textbf{Sarah Rajtmajer} \\
  Information Sciences and Technology, The Pennsylvania State University \\
  \texttt{\{qfm5033, smr48\}@psu.edu}
}


\maketitle
\begin{abstract}
Large language models (LLMs) have emerged as a powerful tool for synthetic data generation. A particularly important use case is producing synthetic replicas of private text, which requires carefully balancing privacy and utility. We propose Realistic and Privacy-Preserving Synthetic Data Generation (RPSG), which uses private seeds and integrates privacy-preserving strategies, including a formal differential privacy (DP) mechanism in the candidate selection, to generate realistic synthetic data. Comprehensive experiments against state-of-the-art private synthetic data generation methods demonstrate that RPSG achieves high fidelity to private data while providing strong privacy protection.  \footnote{All code is available at \url{https://github.com/masonmq/rpsg}}
\end{abstract}

\section{Introduction}

Synthetic data generation is an active area of work in natural language processing (NLP) ~\cite{bommasani2019towards,DBLP:conf/iclr/YuNBGI0KLMWYZ22} with applications ranging from clinical text analysis~\cite{DBLP:journals/jamia/WalonoskiKNQMHD18,DBLP:journals/corr/abs-2303-04360} to social media synthesis~\cite{DBLP:journals/corr/abs-2303-04226, DBLP:journals/corr/abs-2302-04062}.

One type of synthetic data of significant practical interest is synthetic replicas of private text ~\cite{DBLP:conf/icml/HouSZCLSFL24,yu2023pubmed}. Oftentimes, text data that would be of benefit, e.g., to researchers, policymakers, or technologists, cannot be shared due to privacy considerations. Text collected from social media platforms, for example, often contains users' voluntarily disclosed personal information--so-called self-disclosures (see~\cite{ashuri2024online} for a recent interdisciplinary review). Such text can be used for user targeting and manipulation; if the personal information shared is more sensitive, e.g., personally identifiable information (PII), the risks can be graver~\cite{DBLP:journals/cbsn/GruzdH18}.

Mainstream methods for generating privacy-preserving synthetic data include fine-tuning LLMs (e.g., DistilGPT2~\cite{DBLP:conf/emnlp/WolfDSCDMCRLFDS20}) with differential privacy (DP) mechanisms applied through extensive modifications of gradient descent during training~\cite{DBLP:conf/icml/HouSZCLSFL24}, as well as using prompt engineering (e.g., GPT-4~\cite{DBLP:journals/corr/abs-2303-08774}) to guide models toward producing semantically similar synthetic data~\cite{DBLP:journals/corr/abs-2406-07217}. However, fine-tuning approaches without rigorous privacy safeguards have been shown to suffer from memorization and leakage vulnerabilities~\cite{DBLP:journals/corr/abs-2205-12506,DBLP:journals/pvldb/LiHXTXHYWHWLHS24}, and are increasingly impractical as many modern LLMs are accessible only via APIs. In contrast, prompt-based methods are model-access agnostic and can be applied to both API-based and open-source models. However, they struggle to generate high-quality synthetic data that preserves the richness and utility of the original private data~\cite{DBLP:conf/icml/Xie0BGYINJZL0Y24}.

\begin{figure*}[t]
    \centering
    \includegraphics[width=0.98\linewidth]{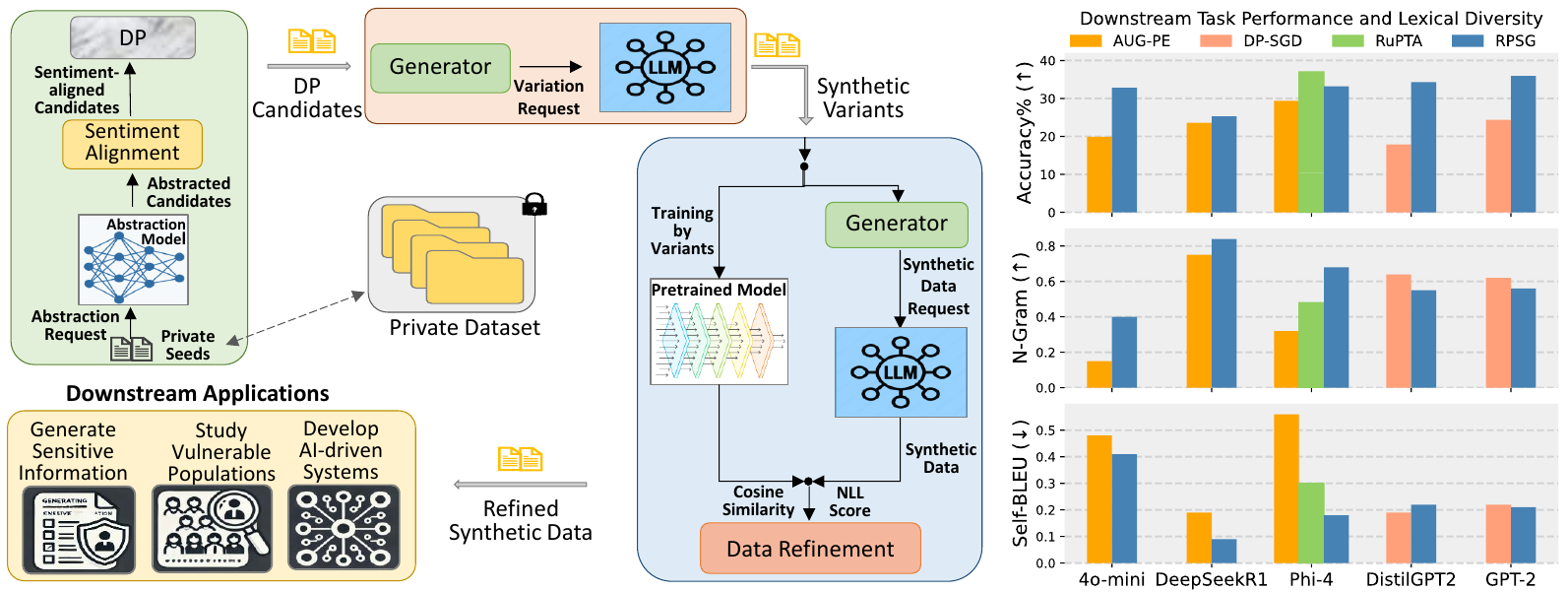}
    \caption{\added{Illustration of the RPSG Method Pipeline.} Comparative performance of RPSG against \added{DP-SGD (\(\epsilon=\infty\)),} AUG-PE (\(\epsilon=\infty\)), and \added{RUPTA} across different LLMs. RPSG achieves higher accuracy, diversity, and lexical quality on several settings, demonstrating advantages over existing approaches.}
    \label{fig:RPSG}
\end{figure*}

In response to these limitations, we propose a novel method for realistic and privacy-preserving synthetic data generation (RPSG) (Figure~\ref{fig:RPSG}). RPSG is designed using private data as seeds to generate high-quality synthetic data that closely resembles the original while reducing privacy risk. In Phase 1, an abstraction model produces multiple sentiment-aligned abstracted candidates for each private seed, reducing identifiable patterns and semantic structures; a formal DP mechanism is then applied to select candidates. In Phase 2, an LLM generates variations of these DP-selected candidates. In Phase 3, synthetic variants are refined to reduce memorization risks and redact any PII. The resulting samples constitute a final set of realistic and privacy-preserving synthetic data, suitable for downstream applications.

We conduct comprehensive experiments to evaluate our method on a benchmark dataset from PubMed~\cite{yu2023pubmed} and an original dataset created from Reddit (details provided in~\S\ref{subsec:sd_dataset}). Our approach is compared against \added{three} baselines: \added{a gradient-based method, DP-SGD~\cite{DBLP:conf/acl/YueILKMS0LS23},} a prompt-based method, AUG-PE~\cite{DBLP:conf/icml/Xie0BGYINJZL0Y24}, \added{and a one-to-one rewriting method, RUPTA~\cite{DBLP:conf/acl/0004ZG25}.} We evaluate \added{under varying privacy budgets} across multiple dimensions, including downstream task performance, \added{sentiment alignment,} lexical diversity, semantic and distributional similarity, and resistance to membership inference attacks (MIAs)~\cite{DBLP:conf/uss/CarliniTWJHLRBS21,DBLP:conf/sp/CarliniCN0TT22,DBLP:conf/acl/MatternMJSSB23}. We further assess protection against PII leakage~\cite{DBLP:conf/nips/WangCPXKZXXDSTA23}, qualitative and structural aspects, computational efficiency, and conduct ablation studies. Our contributions are:
\begin{itemize}

\item We propose RPSG, a novel method for privacy-preserving synthetic data generation. RPSG directly uses private data as seeds to produce one-to-one mapped synthetic text, integrating a formal DP candidate selection mechanism and additional empirical privacy safeguards while achieving improved utility.

\item We evaluate RPSG against \added{gradient-based and} prompt-based DP baselines, \added{as well as a one-to-one rewriting baseline} under \added{varying privacy budgets,} observing competitive utility, diversity, privacy, and efficiency.

\item We demonstrate that RPSG provides strong empirical privacy protection, achieving stronger resistance to MIAs (AUC $\approx$ 50\%) compared to baselines. We also identify common methodological pitfalls highlighted in recent critiques, supporting the rigor of our privacy evaluation.

\end{itemize} 

The remainder of this paper is organized as follows. We first overview related work, followed by our methodology. We then describe the experimental methodology. Subsequent sections detail results and conclusions. Appendices provide additional results and detailed experimental settings.

\section{Related Work}

\subsection{DP Fine-Tuning Methods and Limitations}

With the growing interest in DP for AI applications~\cite{DBLP:conf/focs/BassilyST14,DBLP:conf/iclr/PapernotAEGT17,DBLP:conf/iclr/PapernotSMRTE18}, a prominent approach within this space is training models while ensuring formal privacy guarantees~\cite{DBLP:conf/cvpr/00050CW20,DBLP:conf/csfw/Mironov17}. In particular, DP synthetic data generation has proposed various methods to produce useful data under privacy constraints~\cite{DBLP:conf/iclr/HoltzmanBDFC20}. Among these, fine-tuning pre-trained LLMs using DP-SGD~\cite{DBLP:conf/ccs/AbadiCGMMT016,DBLP:conf/acl/YueILKMS0LS23} has emerged. It enforces DP during model training by incorporating gradient clipping and noise addition into the optimization process~\cite{DBLP:conf/emnlp/MatternJWSS22,DBLP:journals/corr/abs-2408-07055}.

However, these fine-tuning approaches face several critical limitations. Advanced LLMs such as GPT-5~\cite{openai2025gpt5} and Claude 3.7 Sonnet~\cite{anthropic2025claude37} are only accessible via APIs, making it impossible to apply DP-based fine-tuning directly to them~\cite{DBLP:conf/acl/LongWXZDCW24, DBLP:journals/corr/abs-2410-16811}. Although open-source models, such as Phi-4~\cite{DBLP:journals/corr/abs-2412-08905} and Llama 3.3~\cite{meta2024llama33}, are accessible without APIs, the computational cost of DP fine-tuning is substantial~\cite{DBLP:conf/nips/MalladiGNDL0A23,kurakin2023harnessing}. In addition, applying private data directly for fine-tuning raises privacy risks due to model memorization \cite{DBLP:journals/corr/abs-2409-11423,DBLP:conf/uss/CarliniTWJHLRBS21,DBLP:journals/corr/abs-2404-01231}.

\subsection{Prompt-Based Methods and Limitations}
Prompt engineering has emerged as another method for synthetic data generation.~\citet{DBLP:journals/corr/abs-2410-12881} provides a pre-trained LLM with a web document and prompts it in a zero-shot manner to generate a conversation.~\citet{DBLP:journals/corr/abs-2410-14208} utilizes synthetic training data to characterize the source model's learning preferences and then train the target model to generate synthetic data.~\citet{DBLP:journals/corr/abs-2410-16534} uses data-driven loss minimization to train a parameterized contextual soft prompt which then is used to steer the frozen LLM to generate synthetic sequences.~\citet{DBLP:conf/icml/Xie0BGYINJZL0Y24} leverages prompts to guide LLMs in generating synthetic data and then compute the similarity to select the best matches. ~\citet{DBLP:journals/corr/abs-2410-18588} fine-tunes a student model using shorter, less expensive vanilla prompts to generate final synthetic data. By leveraging prompting instead of fine-tuning, these methods avoid exposing sensitive private data to the LLMs. However, the synthetic data produced through prompt engineering often fails to adequately mimic private data~\cite{DBLP:conf/acl/LongWXZDCW24}, limiting its applicability as a substitute for private data.

\subsection{\added{One-to-one Rewriting Methods}}
\label{subsec:1:1_rewriting}
\added{Another direction of research adopts one-to-one rewriting strategies, which rewrite each private text into exactly one synthetic text. The rewriting process is guided by privacy and utility objectives, so that the transformed text conceals sensitive attributes while retaining usefulness for downstream applications.~\citet{DBLP:conf/acl/0004ZG25} propose an anonymization framework in which a privacy evaluator, a utility evaluator, and an optimizer interact to refine each text until the stopping criteria are met. ~\citet{DBLP:journals/corr/abs-2407-02956} address attribute inference by rewriting texts so that an adversary is misled toward a chosen incorrect attribute value. Their method employs an iterative loop in which an adversarial evaluator predicts attributes and provides explanations, and an anonymizer generates revised candidates until the adversary no longer recovers the true value.}  

\added{These one-to-one rewriting approaches demonstrate the potential of rewriting-based anonymization for privacy preservation. }

\subsection{Privacy Risks and Evaluation Methods}

One reason for creating synthetic data is to protect sensitive information, addressing situations where original data cannot be shared due to legal or ethical constraints. Therefore, approaches to synthetic data generation must consider and minimize privacy leakage. However, most synthetic data generation methods prioritize performance metrics~\cite{DBLP:conf/chi/HamalainenTK23}, often overlooking privacy evaluation~\cite{DBLP:journals/corr/abs-2306-01684}. Among privacy-focused approaches,~\citet{DBLP:conf/acl/DouKNKDRX24} study online self-disclosure detection and abstraction, but their privacy analysis focuses on disclosure abstraction and user judgment.~\citet{DBLP:journals/corr/abs-2406-07217} measures personal attribute inference risks but fails to address membership inference risks~\cite{DBLP:conf/sp/ShokriSSS17}. While~\citet{DBLP:conf/icml/Xie0BGYINJZL0Y24} employs MIAs to evaluate synthetic data, its approach generates data directly from prompts, making the MIAs less relevant to actual privacy risks since the synthetic samples are derived from prompt-based generation instead of being grounded in private data.

\section{RPSG Algorithm}
The RPSG algorithm (see Algorithm~\ref{alg:main}) takes as input: the private dataset, $\mathcal{D}_{\textrm{pri}}$; the  abstraction model, $\Psi$; the pretrained model, $\mathcal{M}$; the number of private seeds, $N$; the size of private dataset, $N_{\text{priv}}$; and Negative Log-Likelihood (NLL) percentile $\alpha$. The output is the synthetic dataset, $\mathcal{S}_{\textrm{syn}}^{\prime}$. Specifically, it consists of the following modules.

\subsection{Abstraction and Sentiment Alignment}
\label{subsec:data_abstraction}
Using private data for synthetic data generation can improve utility and fidelity, but it also increases the risk of exposing sensitive information. In RPSG, we randomly select $N$ data points from the private dataset $\mathcal{D}_{\textrm{pri}}$, which has a total size of $N_{\text{priv}}$, as seed inputs for synthetic generation, where each private seed is a private sample. However, simple rephrasing of these seeds by LLMs fails to effectively resist MIAs, because such rewording does not significantly disrupt the statistical or semantic predictability inherent in the data, as the high-level meaning and alignment remain largely intact~\cite{DBLP:conf/uss/CarliniTWJHLRBS21}. 

To address this, the PII processor (\textsc{PII\_FILTER}) is first applied to enforce strict regex-based redaction of private seeds, $\mathcal{D}_{\textrm{pri}}^{(N)}$, replacing structured PII-like information with masked tokens (e.g., [MASK]). Let \(\mathcal{D}_{\mathrm{pri}}^{(N)}=\{x^{(1)},\dots,x^{(N)}\}\) denotes the \(N\) private seeds, we then employ an abstraction model, $\Psi$, to transform \added{each private seed \(x\in\mathcal{D}_{\mathrm{pri}}^{(N)}\) into a set of \(m\) abstracted candidates, \(\mathcal{S}_{\mathrm{abc}}(x)=\{s_1,\dots,s_m\}\)}, thereby breaking direct correspondence with the original. \added{Candidate generation is guided not only by semantic similarity but also by sentiment consistency with the private seed, ensuring that the resulting \(m\) candidates preserve expressive intent while mitigating identifiable patterns. Further implementation details are provided in Appendix~\ref{appendix-abstraction-alignment}.}

\begin{algorithm}[H]
\small
\caption{RPSG Method}
\label{alg:main}

\textbf{Input}: private dataset $\mathcal{D}_{\textrm{pri}}$, text abstraction model $\Psi$, pretrained model $\mathcal{M}$, number of private dataset $N_{\text{priv}}$\\
\textbf{Parameter}: number of private seeds $N$, NLL percentile $\alpha$, privacy budget $\epsilon$, failure probability $\delta$\\
\textbf{Output}: synthetic dataset $\mathcal{S}_{\textrm{syn}}^{\prime}$\\
\begin{algorithmic}[1] 
\STATE $\mathcal{S}_{\textrm{abc}} \gets \Psi(\textsc{PII\_FILTER}(\mathcal{D}_{\textrm{pri}}^{(N)}))$
\begingroup
\added{
\STATE $\sigma \leftarrow \frac{\sqrt{2 \cdot \log (1.25 / \delta)} \cdot \Delta u}{\epsilon}$ \;\; (with $\Delta u = 1.0$ and $\delta = 1/(N_{\text{priv}}\log N_{\text{priv}})$)
\FOR {$x \in \mathcal{D}_{\mathrm{pri}}^{(N)}$}
    \FOR {$s_j \in \mathcal{S}_{\mathrm{abc}}(x)$}
        \STATE $u_j \gets u(x, s_j)$
        \STATE $\tilde{u}_j \gets u_j + \eta_j$, \quad where $\eta_j \sim \mathcal{N}(0, \sigma^2)$ independently across $j$
    \ENDFOR
    \STATE $s_{\mathrm{dpc}}(x) \gets \arg\max_{s_j \in \mathcal{S}_{\mathrm{abc}}(x)} \tilde{u}_j$
\ENDFOR
\STATE $\mathcal{S}_{\textrm{var}} \gets \textsc{SYN\_GEN}(\mathcal{S}_{\textrm{dpc}})$, where \(s_{\mathrm{dpc}}(x) \in \mathcal{S}_{\mathrm{dpc}}\)
}
\endgroup
\STATE $f_{\phi} \gets \arg\min_{f} \sum_{s_{\textrm{var},i} \in \mathcal{S}_{\textrm{var}}} \mathcal{L}(f, s_{\textrm{var},i})$
\STATE $\mathcal{S}_{\textrm{syn}} \gets \textsc{SYN\_GEN}(\mathcal{S}_{\textrm{var}})$
\STATE $\mathcal{S}_{\textrm{syn}}^{\prime} \gets \textsc{COS\_FILTER}(f_{\phi}, \mathcal{S}_{\textrm{syn}}, \mathcal{D}_{\textrm{pri}}^{(N)})$
\FOR{{$s_{\textrm{syn},i}^{\prime} \in \mathcal{S}_{\textrm{syn}}^{\prime}$}}    
    \STATE $\mathcal{L}(f_{\phi}, s_{\textrm{syn},i}^{\prime}) = -\frac{1}{T} \sum_{t=1}^{T} \log \mathcal{P}_{\phi}(w_t \mid w_{<t})$, where $T$ denotes the total number of tokens in $s_{\textrm{syn},i}^{\prime}$
\ENDFOR
\STATE $\tau \gets \text{percentile}_{\alpha}(\{\mathcal{L}(f_{\phi}, s_{\textrm{syn},i}^{\prime})\})$
\STATE $\mathcal{S}_{\textrm{syn}}^{\prime} \gets \{ \textsc{PII\_FILTER}(s_{\textrm{syn},i}^{\prime}) \mid \mathcal{L}(f_{\phi}, s_{\textrm{syn},i}^{\prime}) > \tau \}$

\STATE \textbf{return} $\mathcal{S}_{\textrm{syn}}^{\prime}$
\end{algorithmic}
\end{algorithm}

\subsection{\added{Formal DP in Candidate Selection}}
\added{\label{subsec:dp_selection}}
\added{To select a final DP-protected candidate from \(\mathcal{S}_{\mathrm{abc}}(x)\) with formal privacy guarantees, we apply the Gaussian mechanism~\cite{DBLP:conf/csfw/Mironov17} to the candidate selection process. Formal definitions of DP and the Gaussian mechanism are provided in Appendix~\ref{appendix-preliminaries}. }

\added{Specifically, we define a bounded utility function \(u(x,s_j)\in[0,1]\) that measures the semantic similarity between the private seed \(x\) and a candidate \(s_j\), instantiated as normalized cosine similarity in our implementation. Let \(u_j:=u(x,s_j)\) for \(j=1,\dots,m\). } 

\added{Given a privacy budget \(\epsilon\) and failure probability \(
\delta = \frac{1}{N_{\text{priv}} \cdot \log N_{\text{priv}}}
\)
~\cite{DBLP:conf/acl/YueILKMS0LS23}, we compute the standard deviation \(\sigma\) of the Gaussian noise as: }

\begingroup
\added{
\begin{equation} 
\small
\sigma = \frac{\sqrt{2 \cdot \log \left( \frac{1.25}{\delta} \right)} \cdot \Delta u}{\epsilon}
\label{eq:gaussian_sigma}
\end{equation}
}
\endgroup
\added{where \(\Delta u = 1.0\) is the \(\ell_2\)-sensitivity of the utility function~\cite{DBLP:journals/fttcs/DworkR14}, since \(u(x,s_j) \in [0,1]\).}

\added{For each private seed \(x \in \mathcal{D}_{\mathrm{pri}}^{(N)}\) and each candidate \(s_j \in \mathcal{S}_{\mathrm{abc}}(x)\), we generate a noisy score 
\(
\tilde{u}_j = u_j + \eta_j, \quad \eta_j \sim \mathcal{N}(0, \sigma^2)
\). The final DP-protected candidate for seed \(x\) is denoted \(s_{\mathrm{dpc}}(x)\) and is selected as:}

\begingroup
\added{
\begin{equation}
\small
s_{\mathrm{dpc}}(x) = \arg\max_{s_j \in \mathcal{S}_{\mathrm{abc}}(x)} \tilde{u}_j
\label{eq:dp_selection}
\end{equation}
}
\endgroup
\added{where \(s_{\mathrm{dpc}}(x) \in \mathcal{S}_{\mathrm{dpc}}\).}

\added{This ensures that the selection process satisfies \((\epsilon, \delta)\)-DP, while still favoring candidates with high sentiment and semantic similarity to the private seed. A formal statement of the DP guarantee for candidate selection step is provided in Appendix~\ref{appendix-preliminaries-privacy-guarantee}.
}

\subsection{Synthetic Data Generation}
\label{subsec:synthetic_data_generation}
The synthetic data generation module (\textsc{SYN\_GEN}) uses an LLM to generate synthetic data via prompting and is applied twice. First, \added{DP-protected} candidates, $\mathcal{S}_{\textrm{dpc}}$, are input to produce synthetic variants $\mathcal{S}_{\textrm{var}}$. Second, qualified variants are used to create the synthetic data, $\mathcal{S}_{\textrm{syn}}$, for further refinement. Prompts are designed to encourage the LLM to generate outputs that reflect the structure, semantics, and diversity of the original input. Prompt design is further detailed in Appendix~\ref{appendix-prompt-design}. 

\subsection{Data Refinement}
\label{subsec:data_refinement}
MIAs exploit the observation that models tend to memorize training samples when they assign consistently low token-wise NLL scores~\cite{DBLP:conf/acl/MatternMJSSB23}, or high confidence to specific samples and their slight variations. Stability in NLL scores thus provides a strong signal for membership inference, particularly in overfitted models. The \textsc{REFINEMENT} procedure leverages this property to remove synthetic samples exhibiting memorization, reducing susceptibility to MIAs.

\textbf{Step 1:} We begin by using the synthetic variants, $\mathcal{S}_{\textrm{var}}$, to fine-tune a pretrained model $\mathcal{M}$, resulting in a surrogate model, $f_{\phi}$. This surrogate model captures the distributional properties of $\mathcal{S}_{\textrm{var}}$ and reflects the memorization risk associated with overfitting to it. 

\begin{table*}
    \centering  
    \makebox[0.98\textwidth][c]{
    \begin{minipage}{0.48\textwidth}
    { 
    \fontsize{6.5}{5.5}\selectfont
    \begin{tabular}{>{\centering\arraybackslash}p{0.7cm} >{\centering\arraybackslash}p{1cm} >{\centering\arraybackslash}p{0.9cm} | >{\centering\arraybackslash}p{0.8cm} | >{\centering\arraybackslash}p{0.6cm} | >{\centering\arraybackslash}p{0.6cm}}
        \toprule

        \textbf{Dataset} & \textbf{Model} & \textbf{Method} & \textbf{Acc}(\%)($\uparrow$) & \textbf{Loss}($\downarrow$) & \textbf{PPL}($\downarrow$) \\    
        \midrule       
        \multirow{16}{*}{Reddit} & \multirow{2}{*}{GPT-3.5} & AUG-PE  & 31.7 &3.17& 23.9 \\
         &  & RPSG & \textbf{34.0} & \textbf{3.16} & \textbf{23.5} \\
        \cmidrule{2-6}
         & \multirow{2}{*}{GPT-4o-mini} & AUG-PE  & 19.9 & 4.01 & 55.1 \\
         &  & RPSG & \textbf{32.8} & \textbf{3.23} & \textbf{25.2} \\
         \cmidrule{2-6}
         & \multirow{2}{*}{DeepSeek-R1} & AUG-PE  & 23.6 & 3.93 & 50.9 \\
         &  & RPSG & \textbf{25.3} & \textbf{3.76} & \textbf{42.9} \\
        \cmidrule{2-6}
         & \multirow{2}{*}{Phi-4} & AUG-PE  & 29.4 & 3.50 & 33.1 \\
         &  & RPSG & \textbf{33.2} & \textbf{3.15} & \textbf{23.3} \\

        \cmidrule[\heavyrulewidth]{2-6}

         & \multirow{2}{*}{DistilGPT2} &  DP-SGD  
        &  17.9 &  4.51 &  91.0 \\
         &  &  RPSG 
        & \textbf{34.3} & \textbf{3.18} & \textbf{24.0} \\

        \cmidrule{2-6}

        & \multirow{2}{*}{ GPT-2} &  DP-SGD  
        &  24.4 &  3.70 &  40.4 \\
         &  &  RPSG 
        & \textbf{35.9} & \textbf{3.06} & \textbf{21.1} \\
         
        \midrule \midrule
        \multirow{10}{*}{PubMed} & \multirow{2}{*}{GPT-3.5} & AUG-PE  & 34.4 & 3.04 & 20.9 \\
         &  & RPSG & \textbf{34.4} & 3.13 & 22.9 \\
        \cmidrule{2-6}
         & \multirow{2}{*}{GPT-4o-mini} & AUG-PE  & 35.8 & 2.96 & 19.3 \\
         &  & RPSG & \textbf{36.1} & 3.11 & 22.5 \\
         \cmidrule{2-6}
         & \multirow{2}{*}{DeepSeek-R1} & AUG-PE  & 10.3 & 5.17 & 179 \\
         &  & RPSG & \textbf{13.1} & \textbf{4.66} & \textbf{105} \\
        \cmidrule{2-6}
         & \multirow{2}{*}{Phi-4} & AUG-PE  & 32.4 & 3.41 & 30.0 \\
         &  & RPSG & \textbf{32.9} & \textbf{3.34} & \textbf{28.1} \\
        \bottomrule
    \end{tabular}
    \caption{Language Modeling Performance Results\\for the non-DP Baseline (\(\epsilon=\infty\)).}
    \label{tb:downstream_eval}
    }   
    \end{minipage}\hfill
    
    \begin{minipage}{0.48\textwidth}
    \centering
    { 
    \fontsize{6.5}{5.5}\selectfont 
   \begin{tabular}{>{\centering\arraybackslash}p{0.6cm} >{\centering\arraybackslash}p{0.9cm} >{\centering\arraybackslash}p{0.9cm} | >{\centering\arraybackslash}p{1.4cm} | >{\centering\arraybackslash}p{1.2cm}}
        \toprule
        \textbf{Dataset} & \textbf{Model} & \textbf{Method} & \textbf{Self-BLEU}($\downarrow$) & \textbf{N-Gram}($\uparrow$) \\
        \midrule
        \multirow{16}{*}{Reddit} 
         & \multirow{2}{*}{GPT-3.5} 
         & AUG-PE  & 0.61 & 0.11 \\
         &  & RPSG & \textbf{0.34} & \textbf{0.51} \\
        \cmidrule{2-5}
         & \multirow{2}{*}{GPT-4o-mini} 
         & AUG-PE  & 0.48 & 0.15 \\
         &  & RPSG & \textbf{0.41} & \textbf{0.40} \\
        \cmidrule{2-5}
         & \multirow{2}{*}{DeepSeek-R1} 
         & AUG-PE  & 0.19 & 0.75 \\
         &  & RPSG & \textbf{0.09} & \textbf{0.84} \\
        \cmidrule{2-5}
         & \multirow{2}{*}{Phi-4} 
         & AUG-PE  & 0.56 & 0.32 \\
         &  & RPSG & \textbf{0.18} & \textbf{0.68} \\

        \cmidrule[\heavyrulewidth]{2-5}
         & \multirow{2}{*}{DistilGPT2} 
         &  DP-SGD  &  0.19 &  0.64 \\
         &  &  RPSG &  0.22 &  0.55 \\
        \cmidrule{2-5}
         & \multirow{2}{*}{GPT-2} 
         &  DP-SGD  &  0.22 &  0.62 \\
         &  &  RPSG & \textbf{0.21} &  0.56 \\

        \midrule \midrule
        \multirow{10}{*}{PubMed} 
         & \multirow{2}{*}{GPT-3.5} 
         & AUG-PE  & 0.62 & 0.34 \\
         &  & RPSG & \textbf{0.29} & \textbf{0.57} \\
        \cmidrule{2-5}
         & \multirow{2}{*}{GPT-4o-mini} 
         & AUG-PE & 0.72 & 0.28 \\
         &  & RPSG & \textbf{0.33} & \textbf{0.53} \\
        \cmidrule{2-5}
         & \multirow{2}{*}{DeepSeek-R1} 
         & AUG-PE  & 0.19 & 0.74 \\
         &  & RPSG & 0.20 & \textbf{0.81} \\
        \cmidrule{2-5}
         & \multirow{2}{*}{Phi-4} 
         & AUG-PE  & 0.51 & 0.39 \\
         &  & RPSG & \textbf{0.27} & \textbf{0.63} \\
        \bottomrule
    \end{tabular}
    \caption{Lexical Diversity Results for the non-DP Baseline (\(\epsilon=\infty\)).}
    \label{tb:diversity_eval}
    }
    \end{minipage}
    }
\end{table*}

\textbf{Step 2:} To effectively identify data with highest similarity to the private data, we utilize the embeddings from $f_{\phi}$ and compute cosine similarity scores between samples from newly generated $\mathcal{S}_{\textrm{syn}}$ and $\mathcal{D}_{\textrm{pri}}^{(N)}$. Synthetic samples exhibiting the highest similarity are identified and removed, and the retained samples form $\mathcal{S}_{\textrm{syn}}^{\prime}$.

\textbf{Step 3:} For each synthetic sample, $s_{\textrm{syn},i}^{\prime}$, remaining after cosine similarity filtering, we compute its NLL score. This score reflects how confidently the surrogate model predicts the text, and thus serves as a proxy for memorization risk, given as:
\begin{equation}
\small
\mathcal{L}(f_{\phi}, s) = -\frac{1}{T}\sum_{t=1}^{T}\log \mathcal{P}_{\phi}(w_t \mid w_{<t})
\label{eq:nll_private_syn}
\end{equation}
where $s$ represents a sample of $s_{\textrm{syn},i}^{\prime}$, $w_t$ is the $t$-th token in $s$, and $\mathcal{P}_{\phi}(w_t\mid w_{<t})$ is the conditional probability assigned by surrogate model $f_{\phi}$ given the preceding context. Here, $T$ denotes the total number of tokens in $s$.

\textbf{Step 4:} To mitigate memorization, we retain only those synthetic samples whose NLL scores are above a threshold $\tau$, indicating they are less likely to be memorized. The threshold is defined as the $\alpha$-percentile of the NLL score distribution:
\begin{equation}
\small
\tau = \text{Percentile}_{\alpha} \left( \left\{ \mathcal{L}(f_\phi, s^{\prime}) \right\}_{s^{\prime} \in \mathcal{S}_{\textrm{syn}}^{\prime}} \right)
\label{eq:percentile_tau}
\end{equation}
Experimental details regarding the selection of the cosine similarity and $\alpha$-percentile thresholds can be found in Appendix~\ref{appendix-synthetic-data-generation-configuration}.

Finally, the PII processor is applied to this dataset to further mitigate privacy risks, yielding our final robust synthetic dataset. 

\section{Experimental Methodology}
\subsection{Datasets}

\subsubsection{PubMed Dataset}
The PubMed abstracts corpus is widely used as a benchmark for language modeling and fine-tuning (e.g., \cite{DBLP:journals/health/GuTCLULNGP22}). In our experiments, we use the subset of abstracts published between August 1 and August 7, 2023, crawled by \citet{yu2023pubmed}, comprising 75,329 training, 14,423 validation, and 4,453 test samples.

\subsubsection{Reddit Dataset Construction}
\label{subsec:sd_dataset}
The availability of standardized benchmark datasets for studying NLP tasks and evaluating LLMs remains limited~\cite{DBLP:journals/corr/abs-2404-06001}. We construct a novel social media dataset focused on English-language conversations on Reddit associated with financial hardship and poverty. These conversations generally contain a rich amount of self-disclosures---personal information related to finance, health, family, age, location, and similar. Prior research indicates that Reddit users tend to share more sensitive and detailed personal information than on other platforms~\cite{DBLP:conf/icwsm/ChoudhuryD14,du2024toward}.

Specifically, we manually selected subreddits associated with financial hardship and poverty, and applied a keyword filtering strategy designed to capture specific language related to economic challenges (details on subreddit selection and the full list of keywords are provided in Appendix~\ref{appendix-dataset-construction}). We collected posts published between January 1, 2024, and March 31, 2025, ensuring that the majority of data falls after the training cutoffs of both GPT-4 (October 2023) and Phi-4 (June 2024). This timing minimizes the likelihood that these LLMs were exposed to our dataset during training, supporting its value for evaluating model generalization and privacy behavior. The final dataset consists of 8,948, 1,000, and 1,000 posts in the training, validation, and test sets, respectively. All posts were publicly available and collected in accordance with Reddit’s terms of service. 

\subsection{Models}
Facebook/bart-large-cnn~\cite{DBLP:conf/acl/LewisLGGMLSZ20} served as the abstraction model, and \added{siebert/sentiment-roberta-large-english~\cite{hartmann2023} functioned as the sentiment classification model} (\S\ref{subsec:data_abstraction}). \added{DistilGPT2~\cite{sanh2019distilbert}, GPT-2}, GPT-3.5-turbo, GPT-4o-mini~\cite{DBLP:journals/corr/abs-2303-08774}, DeepSeek-R1~\cite{DBLP:journals/corr/abs-2501-12948}, Phi-4-mini, and Phi-4~\cite{DBLP:journals/corr/abs-2412-08905} were leveraged as LLMs for synthetic data generation (\S\ref{subsec:synthetic_data_generation}). BERT-small~\cite{DBLP:journals/corr/abs-1908-08962} was employed as the pretrained base model to produce the surrogate model by fine-tuning (\S\ref{subsec:data_refinement}), and also serves as the downstream model to evaluate performance (\S\ref{subsubsec:down_task_performance}). The sentence transformer sentence-t5-base~\cite{DBLP:conf/emnlp/ReimersG19} was used as the embedding model to calculate the semantic similarity (\S\ref{subsubsec:distri_semantic_similarity}). Finally, bigcode/starpii~\cite{DBLP:journals/corr/abs-2301-03988} serves as the detection model to evaluate PII leakage (\S\ref{subsubsec:pii_leakage}).

\subsection{Metrics}
\label{subsec:metrics}
\added{To address potential blind spots in individual metrics~\cite{DBLP:conf/acl/HeZ0KCGT23} and ensure a comprehensive evaluation,} we employ a diverse set of complementary metrics.

\subsubsection{Performance Evaluation Metrics}
We evaluate synthetic data along four dimensions: (1) language modeling performance; \added{(2) sentiment alignment;} (3) lexical diversity; and (4) distributional and semantic similarity to the private data. For language modeling performance, we fine-tune BERT-small on the synthetic text and evaluate Next-word Prediction Accuracy and Perplexity~\cite{DBLP:conf/icml/Xie0BGYINJZL0Y24}. \added{For sentiment alignment, we predict sentiment on each private seed and its synthetic counterpart and report the Sentiment Alignment, following the standard hit rate evaluation in ~\cite{hartmann2023}, which supports evaluating sentiment faithfulness of synthetic counterpart to private seeds. This evaluation is conducted only on the Reddit dataset, as PubMed samples do not contain sentiment-related attributes.} To measure lexical diversity, we use Self-BLEU~\cite{DBLP:conf/sigir/ZhuLZGZWY18} and n-gram diversity~\cite{DBLP:journals/corr/abs-1904-03971}. For distributional and semantic similarity, we employ Fréchet Inception Distance (FID), Precision, Recall, F1, Mauve, Kullback–Leibler Divergence (KLD), Total Variation Divergence (TVD), Wasserstein Metric Distance (WMD), and Sinkhorn Loss (SL) to assess embedding-level alignment with private data~\cite{DBLP:conf/icml/Xie0BGYINJZL0Y24}. 

\subsubsection{Privacy Evaluation Metrics}
The privacy of synthetic data is evaluated across two key dimensions: (1) resistance to MIAs; and (2) protection against PII leakage. To evaluate resistance to MIAs, we compute the Area Under the Curve (AUC) scores across three standard attacks: threshold-based perplexity (PPL)~\cite{DBLP:conf/uss/CarliniTWJHLRBS21}; log-perplexity ratio against a reference model (REFER)~\cite{DBLP:conf/uss/CarliniTWJHLRBS21}; and likelihood ratio (LIRA)~\cite{DBLP:conf/sp/CarliniCN0TT22}. We follow the evaluation methodology introduced by~\citet{DBLP:conf/acl/MatternMJSSB23} to assess the effectiveness of synthetic data in mitigating MIA risks. To assess PII leakage, we use the PII successful extraction~\cite{DBLP:conf/nips/WangCPXKZXXDSTA23}, which quantifies the extent to which PII-like content is retained in the generated data.

\subsection{Baselines}
\label{subsec:baselines}
We compare RPSG against \added{DP-SGD~\cite{DBLP:conf/acl/YueILKMS0LS23},} AUG-PE~\cite{DBLP:conf/icml/Xie0BGYINJZL0Y24}, \added{and RUPTA~\cite{DBLP:conf/acl/0004ZG25},} which represent three different approaches to privacy-preserving text generation. \added{DP-SGD uses gradient-based training with DP,} AUG-PE reflects the prompt-based line of work, \added{and RUPTA is a one-to-one rewriting method that rewrites each instance for anonymization. Taken together, these baselines provide a broad and balanced context for evaluating RPSG.}

\begin{table*}
    \centering
    {
    \fontsize{7.5}{8.5}\selectfont  
    \begin{tabular}{>{\centering\arraybackslash}m{0.6cm} 
                    >{\centering\arraybackslash}m{0.6cm} 
                    >{\centering\arraybackslash}m{0.8cm} |
                    >{\centering\arraybackslash}m{0.9cm} 
                    >{\centering\arraybackslash}m{0.8cm} 
                    >{\centering\arraybackslash}m{1.1cm} 
                    >{\centering\arraybackslash}m{1.3cm} 
                    >{\centering\arraybackslash}m{1.3cm} |
                    >{\centering\arraybackslash}m{0.3cm} 
                    >{\centering\arraybackslash}m{0.6cm} 
                    >{\centering\arraybackslash}m{0.5cm}}
    
        \toprule
        \multirow{3}{*}{\textbf{Dataset}} & 
        \multirow{3}{*}{\textbf{Model}} & 
        \multirow{3}{*}{\textbf{Method}} & 
        \multirow{3}{*}{\textbf{Acc(\%)($\uparrow$)}} &
        \multirow{3}{*}{\textbf{PPL($\downarrow$)}} &
        \multirow{2}{*}{\textbf{Sentiment}} &
        \multirow{3}{*}{\textbf{Self-BLEU($\downarrow$)}} &
        \multirow{3}{*}{\textbf{N-Gram($\uparrow$)}} &
        \multicolumn{3}{c}{\textbf{AUC}} \\
        \cmidrule{9-11}
         &  &  &  &  & \textbf{Align(\%)($\uparrow$)} &  &  & 
         \textbf{PPL} & \textbf{REFER} & \textbf{LIRA} \\
        \midrule
        \multirow{2}{*}{Reddit} 
         & \multirow{2}{*}{Phi-4} 
         & RUPTA & 37.0 & 17.4 & 90.8 & 0.30 & 0.48 & 78.4 & 28.6 & 64.0 \\
         &  & RPSG & 33.2 & 23.3 & \textbf{92.1} & \textbf{0.18} & \textbf{0.68} & \textbf{54.1} & \textbf{50.9} & \textbf{50.0} \\
        \bottomrule
    \end{tabular}
    \caption{ Comparison of RPSG and RUPTA under Phi-4 across Utility, Diversity, Sentiment, and Privacy Metrics.}
    \label{tb:rpsg_vs_rupta_phi4}
    }
\end{table*}

\section{Results}

For AUG-PE, we report results with GPT-3.5, GPT-4o-mini, Phi-4, and DeepSeek-R1; \added{for DP-SGD, with DistilGPT2 and GPT-2; and for RuPTA, with Phi-4. Model choices and initial synthetic sample sizes vary by method due to methodological fit and resource constraints: DP-SGD requires local fine-tuning on release-available models, and RUPTA’s iterative rewriting incurs high token budgets on API-based models. Additional details, including sentiment alignment for abstraction (Appendix~\ref{appendix-abstraction-alignment}), privacy alignment (Appendix~\ref{appendix-privacy-alignment}),} and configuration and hyperparameters (\added{Appendix~\ref{appendix-synthetic-data-generation-configuration}}), are provided in the appendices. 

\subsection{Performance Evaluation}
\label{sub:performance_results}

\subsubsection{Downstream Task Performance} 
\label{subsubsec:down_task_performance}

\paragraph{Language Modeling Performance} 
\label{subsubsec:language_modeling_performance}
Table~\ref{tb:downstream_eval} shows RPSG improves language modeling performance, with larger gains on Reddit (e.g, GPT-4o-mini accuracy 32.8\% vs. 19.9\%, perplexity 25.2 vs. 55.1). On PubMed, gains are smaller but consistent. This suggests RPSG is especially effective when the target data are less familiar to the LLMs. \added{On Reddit, RPSG more than doubles accuracy relative to DP-SGD for DistilGPT2 and clearly outperforms GPT-2. This highlights the advantage of training with prompt-based private synthetic data over direct fine tuning for downstream tasks.}

\added{With finite privacy budgets (Appendix Table~\ref{tb:downstream_eval_w_dp}), RPSG maintains stable accuracy across $\epsilon \in \{4,2,1\}$ on DistilGPT2 and GPT-2, while DP-SGD falls to about five percent across budgets. For GPT-4o-mini and Phi-4, although AUG-PE remains higher under these budgets, RPSG is closer to chance on MIA AUCs (see Appendix Table~\ref{tb:mia_eval_w_dp}), which reflects a privacy utility tradeoff of RPSG.}

\added{Table~\ref{tb:rpsg_vs_rupta_phi4} shows that under Phi-4, RUPTA is higher on accuracy and lower on perplexity, while RPSG remains competitive. The reason is that RPSG targets stronger privacy: AUCs are near chance for RPSG but not for RUPTA, consistent with a privacy utility tradeoff.}

\paragraph{Sentiment Alignment} 
\label{subsubsec:sentiment_align}

\added{Table~\ref{tb:rpsg_vs_rupta_phi4} also shows sentiment alignment as the percentage of private–synthetic pairs whose predicted polarity matches. Under Phi-4 on Reddit, both one-to-one rewriting methods preserve affect well: RPSG attains 92.1\% and RUPTA 90.8\%. This is consistent with the abstraction model fidelity evaluation (\S\ref{sub:abstraction_fidelity}), where abstraction preserves affective polarity.}

\subsubsection{Lexical Diversity}
\label{subsubsec:lexical_diversity}
Table~\ref{tb:diversity_eval} reports Self-BLEU and n-gram diversity (n = 2) at $\epsilon=\infty$. RPSG shows consistently lower Self-BLEU and higher n-gram on both datasets, for example, on Reddit with GPT-4o-mini (Self-BLEU 0.41 vs. 0.48; n-gram 0.40 vs. 0.15). These patterns indicate broader lexical coverage and reduced repetition in RPSG outputs. \added{For DistilGPT2, DP-SGD reports lower Self-BLEU and higher n-gram. Taken together with the utility results, this suggests DP-SGD’s apparent “extra” diversity often reflects noise, while RPSG delivers competitive diversity.}

\added{Appendix Table~\ref{tb:lexical_diversity_w_dp} extends to finite privacy budgets. On Reddit with GPT-4o-mini and Phi-4, RPSG improves diversity as $\epsilon$ tightens and remains above AUG-PE at every budget. DP-SGD shows near zero Self-BLEU and very high n-gram across budgets, together with about five percent accuracy, suggesting noise driven stylistic drift that inflates diversity without adding useful variety.}

\added{Table~\ref{tb:rpsg_vs_rupta_phi4} shows that RPSG attains lower Self-BLEU and higher n-gram, indicating more varied lexical patterns than RUPTA.}

\begin{table}
    \centering
    {
    \fontsize{6.5}{5.5}\selectfont 
    \renewcommand{\arraystretch}{1.1} 
    \begin{tabular}{>{\centering\arraybackslash}m{0.6cm} >{\centering\arraybackslash}m{1.1cm} >{\centering\arraybackslash}m{0.9cm} |>{\centering\arraybackslash}m{0.5cm} | >{\centering\arraybackslash}m{0.7cm} | >
    {\centering\arraybackslash}m{0.6cm}}
    
        \toprule
        \multirow{3}{*}{\textbf{Dataset}} & \multirow{3}{*}{\textbf{Model}} & \multirow{3}{*}{\textbf{Method}} & \multicolumn{3}{c}{\textbf{AUC}} \\
        \cmidrule(lr){4-6}
         &  &  & \textbf{PPL} & \textbf{REFER} & \textbf{LIRA} \\
        \midrule
        \multirow{16}{*}{Reddit} 
         &  \multirow{2}{*}{GPT-3.5} 
         & AUG-PE  & 48.4 & 60.3 & 43.2 \\
         &  & RPSG & 53.9 & \textbf{56.2} & \textbf{44.8} \\
        \cmidrule(lr){2-6}
         & \multirow{2}{*}{\shortstack{GPT-4o-mini}} 
         & AUG-PE & 78.5 & 21.9 & 68.5 \\
         &  & RPSG & \textbf{52.1} & \textbf{59.3} & \textbf{43.7} \\
         \cmidrule(lr){2-6}
         & \multirow{2}{*}{\shortstack{DeepSeek-R1}} 
         & AUG-PE  & 36.1 & 70.1 & 40.5 \\
         &  & RPSG & \textbf{52.1} & \textbf{58.2} & \textbf{45.6} \\
         \cmidrule(lr){2-6}
         & \multirow{2}{*}{\shortstack{Phi-4}} 
         & AUG-PE  & 74.9 & 31.3 & 59.1 \\
         &  & RPSG & \textbf{54.1} & \textbf{50.9} & \textbf{50.0} \\

        \cmidrule[\heavyrulewidth]{2-6}
     
         & \multirow{2}{*}{ DistilGPT2} &  DP-SGD  
        &  44.3 &  54.1 &  50.4 \\
         &  &  RPSG 
        & \textbf{54.0} &  58.4 &  45.2 \\

        \cmidrule{2-6}

        & \multirow{2}{*}{GPT-2} &  DP-SGD  
        &  31.6 &  38.7 &  56.9 \\
         &  &  RPSG 
        & \textbf{54.3} & \textbf{59.1} & \textbf{44.3} \\
         
        \midrule \midrule
        \multirow{10}{*}{PubMed} 
         & \multirow{2}{*}{GPT-3.5} 
         & AUG-PE  & 60.1 & 36.2 & 61.5 \\
         &  & RPSG & \textbf{50.0} & \textbf{42.7} & \textbf{57.7} \\
        \cmidrule(lr){2-6}
         & \multirow{2}{*}{\shortstack{GPT-4o-mini}} 
         & AUG-PE  & 63.1 & 32.6 & 64.7 \\
         &  & RPSG & \textbf{42.0} & \textbf{51.8} & \textbf{52.6} \\
         \cmidrule(lr){2-6}
         & \multirow{2}{*}{\shortstack{DeepSeek-R1}} 
         & AUG-PE  & 51.9 & 55.8 & 45.6 \\
         &  & RPSG & 53.3 & \textbf{44.2} & 55.3 \\
         \cmidrule(lr){2-6}
         & \multirow{2}{*}{\shortstack{Phi-4}} 
         & AUG-PE  & 54.1 & 36.3 & 56.0 \\
         &  & RPSG & 45.0 & \textbf{51.2} & \textbf{50.5} \\
        \bottomrule
    \end{tabular}
    \caption{Evaluation of MIAs for the non-DP Baseline (\(\epsilon=\infty\)): Lower deviation from 50\% indicates stronger privacy.}
    \label{tb:mia_eval}
    }
\end{table}

\subsubsection{Distributional and Semantic Similarity}
\label{subsubsec:distri_semantic_similarity}
\added{Table~\ref{tb:distri_similarity_eval} reports corpus level alignment between synthetic and private data.} Across Reddit and PubMed, RPSG improves global similarity on several key measures relative to AUG-PE, with lower FID and divergence (e.g., Reddit with Phi-4: FID 0.07 vs. 0.19; KLD 1.41 vs. 13.7) and higher Mauve (e.g., PubMed with DeepSeek-R1: 0.87 vs. 0.25). \added{DP-SGD is higher on several corpus similarity metrics, but this must be interpreted with utility in mind. As shown in Table~\ref{tb:downstream_eval}, these runs have much lower downstream accuracy and higher perplexity than RPSG, so corpus alignment and utility can diverge.}

Recall and F1 also improve on several models (e.g., Reddit with GPT-4o-mini: recall 0.33 vs. 0.13, F1 0.47 vs. 0.23), while AUG-PE often shows higher precision and Recall. This can be attributed to the nature of its generation strategy. AUG-PE prompts LLMs with broad, general prompts (e.g., PubMed prompt: "\textit{Please act as a sentence generator for the medical domain. Generated sentences should mimic the style of PubMed journal articles in a professional way or concise manner or creative style or using imagination or in a formal manner.}"; Reddit prompt: "\textit{Using a variety of sentence structures, Write a passage in the tone of a person who is struggling with poverty or broke or homeless or unemployed or unable to afford basic necessities.}") without grounding them in specific seed inputs. As a result, the generated synthetic data tends to follow structurally common and semantically central patterns representative of the overall domain. These generic patterns increase the likelihood that each synthetic sample resembles multiple private samples (leading to higher Precision), and that each private sample retrieves several related synthetic samples (leading to higher Recall). In contrast, RPSG uses a subset of private data as seeds to guide generation. This strategy emphasizes fidelity to specific seed examples over broader coverage of the data distribution. As a result, RPSG’s synthetic samples are less likely to resemble unrelated private data points, inherently leading to lower Precision and Recall in retrieval-based evaluations.

While AUG-PE yields higher Precision and Recall, reflecting how well individual synthetic samples correspond to nearby private samples, RPSG's outputs better capture the overall distributional and semantic properties of the private data. This aligns with RPSG’s stronger performance on global alignment metrics such as FID, KLD, TVD, and Mauve. Thus, the gap in Precision and Recall does not contradict RPSG’s strength in modeling broader characteristics of the private dataset.

\begin{figure}[t]
    \centering
    \begin{minipage}{0.47\linewidth} 
        \centering
        \includegraphics[width=\linewidth,height=3.5cm]{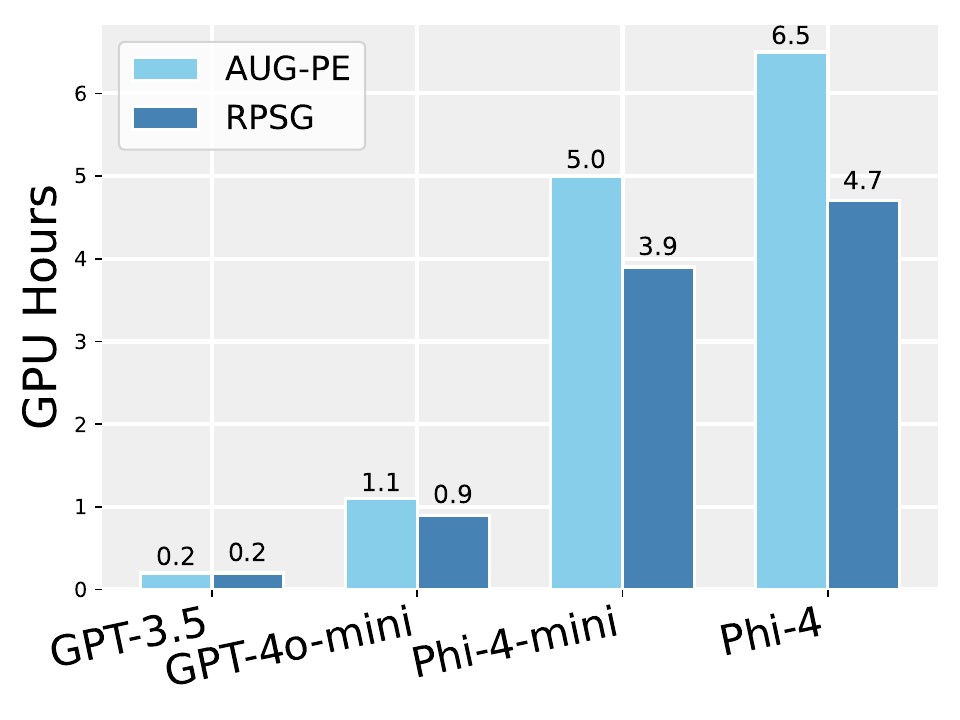}
        \captionsetup{font=footnotesize, labelfont=footnotesize}
        \caption{Efficiency comparison on Reddit for generating 1,000 synthetic samples with no DP (\(\epsilon=\infty\)).}
        \label{fig:eval_gpu}
    \end{minipage}
    \hfill
    \begin{minipage}{0.5\linewidth}
        \centering
        \includegraphics[width=\linewidth,height=3.5cm]{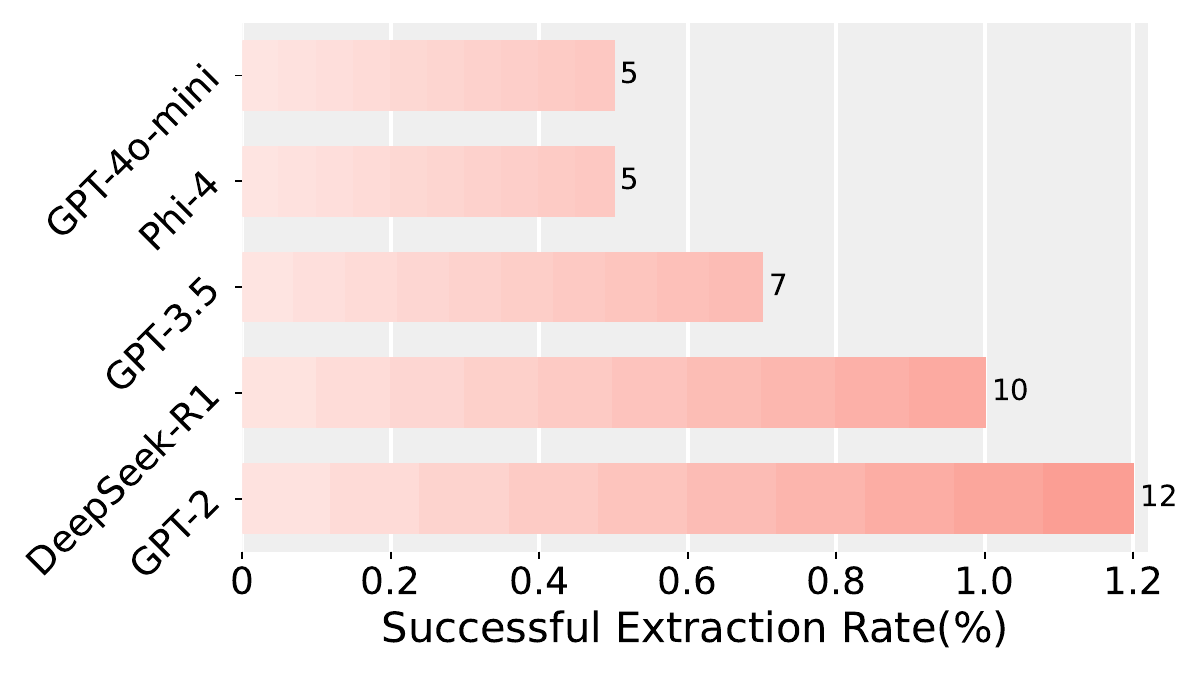}
        \captionsetup{font=footnotesize, labelfont=footnotesize}
        \caption{Evaluation of PII leakage on Reddit across 1,000 synthetic samples with no DP (\(\epsilon=\infty\)).}
        \label{fig:eval_pii}
    \end{minipage}
\end{figure}

\subsection{Privacy Evaluation} 
\label{subsec:privacy_evaluation}
\subsubsection{Resistance to MIAs} 
Table~\ref{tb:mia_eval} reports resistance to MIAs using three AUC scores, which test an adversary’s ability to infer membership of private samples. Lower deviation from 50\% AUC indicates stronger resistance. At $\epsilon=\infty$, RPSG is closer to 50\% than AUG-PE on most LLMs (e.g., on Reddit with GPT-4o-mini, RPSG attains PPL 52.1, REFER 59.3, LIRA 43.7 vs. AUG-PE 78.5, 21.9, 68.5). \added{DP-SGD is sometimes closer (e.g., on DistilGPT2), but at the cost of poor utility. RPSG maintains comparable privacy while dominating in utility (see Table~\ref{tb:downstream_eval}).} 

\added{Table~\ref{tb:rpsg_vs_rupta_phi4} shows that under Phi-4, RPSG is near chance across all three attacks, while RUPTA deviates substantially, indicating stronger resistance for RPSG in this setting.}

\added{Appendix Table~\ref{tb:mia_eval_w_dp} shows that as $\epsilon$ tightens, RPSG often moves REFER and LIRA AUCs toward 50\% on strong LLMs, and is competitive with AUG-PE on these attacks. DP-SGD can be closer on PPL AUC in some cases, yet this coincides with about five percent downstream accuracy (see Table~\ref{tb:downstream_eval_w_dp}), underscoring the need to consider privacy and utility jointly.}

\textbf{\emph{Note.}} A study by~\citet{DBLP:journals/corr/abs-2402-07841} analyzes why MIAs often underperform when applied to LLMs. While we agree with their findings in the context of LLMs, our MIAs evaluation focuses on a different setting. We detail in Appendix~\ref{appendix-for-mias} how our design avoids the methodological pitfalls highlighted in their work.

\subsubsection{Protection against PII Leakage}
\label{subsubsec:pii_leakage}
We assess privacy leakage following the methodology of~\citet{DBLP:conf/nips/WangCPXKZXXDSTA23}, defining it as the rate of successful PII extraction, where a sample is considered leaked if at least one PII entity is detected. We employ their evaluation pipeline and detection model but differ by directly evaluating synthetic data rather than using prompting attacks to elicit memorized data. We do not assume that all detected PII reflects real leakage from the private data. Instead, the PII extraction rate serves as a proxy to measure the likelihood that models generate sensitive-looking content which can still pose privacy risks in deployment.

Figure~\ref{fig:eval_pii} reports PII extraction rates (percentage of samples with detectable PII per 1,000 samples). Our synthetic data demonstrates low privacy leakage (0.5\%–1.2\%). While exact numerical comparisons with~\citet{DBLP:conf/nips/WangCPXKZXXDSTA23} are constrained by methodological differences, our extraction rates are notably lower, indicating minimal privacy risks.

Advanced LLMs (e.g., GPT-4o-mini) exhibit lower extraction rates than older models (e.g., GPT-2), likely due to improved language understanding, instruction-following, and alignment methods, whereas older LLMs more directly replicate input content, increasing potential privacy risks.

\subsection{Sentence-Length Distribution}
\label{appendix-length-distribution-results}
We randomly sampled 1,000 synthetic samples to investigate their length distribution. As illustrated in Figure~\ref{fig:eval_len}, \added{RUPTA reproduces the private distribution almost exactly, since its one-to-one rewriting process preserves the original text length. While this appears ideal for length fidelity, it reflects direct rewriting and therefore provides little additional privacy protection.} RPSG, in contrast, achieves a distribution that closely follows the private data. The peak in the RPSG distribution reflects the preset abstraction length of 150 tokens, with the model typically appending a small amount of additional text to fully express the meaning embedded in the abstracted seeds. AUG-PE, lacking explicit guidance, produces outputs of relatively consistent length, leading to poorer alignment with the private distribution. \added{These comparisons highlight that sentence-length evaluation must be interpreted alongside utility and privacy guarantees, since perfect fidelity to private data does not necessarily imply stronger privacy.}

\subsection{Qualitative Evaluation}
\label{sub:qualitative_structural_eval}

This evaluation includes \added{private--synthetic pair alignment analysis and attribute-level comparison.} These results indicate that RPSG can qualitatively reproduce the characteristics of private data. Full results are presented in \added{Appendix~\ref{appendix-qualitative-evaluation}}.

\subsection{Computational Efficiency Evaluation} 

RPSG demonstrates consistent efficiency gains, achieving speedups ranging from 1.22x to 1.38x. We report GPU hour comparisons for in Figure~\ref{fig:eval_gpu} and Appendix~\ref{appendix-computational-efficiency-results}.

\subsection{\added{Abstraction Fidelity}}
\label{sub:abstraction_fidelity}
\added{To evaluate abstraction fidelity more directly, we measured sentiment alignment between private seeds and their abstracted counterparts generated by the abstraction model. As shown in Appendix Table~\ref{tb:abstraction_eval}, sentiment was preserved at high levels across subsets of 100 samples (86.1\% alignment), 200 samples (89.6\%), and 500 samples (86.5\%). These findings indicate that enforcing sentiment alignment allows abstraction to preserve affective polarity and expressive intent while mitigating identifiability.}

\subsection{Ablation Evaluation}
\label{sub:ablation}
\added{We conduct an ablation experiment (\(\epsilon=\infty\)) on NLL-based filtering, showing its effect on membership inference resistance. This confirms that this setting is necessary for privacy robustness.} We also conduct ablations to examine how sampling temperature and synthetic sample size affect the performance (\(\epsilon=\infty\)). \added{Full details are in Appendix~\ref{appendix-ablation-results}}, with summary visualizations in Appendix Figures~\ref{fig:abla_acc}–\ref{fig:abla_bleu}.

\section{Conclusion}
We have presented RPSG, a realistic and privacy-preserving synthetic data generation method. Empirical results demonstrate that RPSG consistently outperforms baselines in generating high-quality synthetic data, achieving strong utility while safeguarding privacy. This work highlights the opportunity for rigorously integrating privacy into synthetic data generation, and inspires further research and real world adoption of privacy-aware practices.

\newpage

\section*{Limitations}
Despite its effectiveness, RPSG method has two limitations:

\noindent (1) Sensitivity to model-specific characteristics.
Optimal thresholds for similarity-based and NLL-based filtering depend on the generation patterns of the LLMs. Small adjustments to thresholds can impact MIA performance, necessitating careful parameter tuning.

\added{\noindent (2) Aggregation re-profiling risk beyond per sample tests.
Our privacy evaluations focus on per-sample leakage and do not capture an attacker who aggregates many synthetic samples to reconstruct profiles of individuals. Rare attribute combinations repeated across samples could enable re-profiling even when each sample alone appears safe. Measuring and mitigating this corpus-level risk will require attacks and objectives at the set-level, such as linkability tests, membership inference under aggregation, or formal group privacy guarantees. We leave a systematic evaluation of aggregation based attacks and group privacy for future work.}

\section*{Ethical Statement}
Our research involves the collection of publicly available data from Reddit using the official Reddit API. We acknowledge the ethical responsibility to respect user privacy and mitigate any potential risks associated with this data collection. Only publicly available data that is voluntarily shared by Reddit users is used, and we adhere to Reddit's API usage policies to ensure data is gathered in a manner consistent with their terms of service. No attempts were made to de-anonymize users or link the data to individuals beyond what is available through the API. The collected data is used solely for research purposes aimed at understanding and improving synthetic data generation and privacy-preserving methods. The dataset will be shared upon request with verified researchers.

\section*{Acknowledgments}
We gratefully acknowledge the authors of~\citet{DBLP:conf/icml/Xie0BGYINJZL0Y24} for providing their code framework, which we adapted and re-implemented for our work and experiments.

\bibliography{custom}

\begin{thebibliography}{67}
\providecommand{\natexlab}[1]{#1}

\bibitem[{Abadi et~al.(2016)Abadi, Chu, Goodfellow, McMahan, Mironov, Talwar, and Zhang}]{DBLP:conf/ccs/AbadiCGMMT016}
Mart{\'{\i}}n Abadi, Andy Chu, Ian~J. Goodfellow, H.~Brendan McMahan, Ilya Mironov, Kunal Talwar, and Li~Zhang. 2016.
\newblock \href {https://doi.org/10.1145/2976749.2978318} {Deep learning with differential privacy}.
\newblock In \emph{Proceedings of the 2016 {ACM} {SIGSAC} Conference on Computer and Communications Security, Vienna, Austria, October 24-28, 2016}, pages 308--318. {ACM}.

\bibitem[{Abdin et~al.(2024)Abdin, Aneja, Behl, Bubeck, Eldan, Gunasekar, Harrison, Hewett, Javaheripi, Kauffmann, Lee, Lee, Li, Liu, Mendes, Nguyen, Price, de~Rosa, Saarikivi, Salim, Shah, Wang, Ward, Wu, Yu, Zhang, and Zhang}]{DBLP:journals/corr/abs-2412-08905}
Marah~I Abdin, Jyoti Aneja, Harkirat~S. Behl, S{\'{e}}bastien Bubeck, Ronen Eldan, Suriya Gunasekar, Michael Harrison, Russell~J. Hewett, Mojan Javaheripi, Piero Kauffmann, James~R. Lee, Yin~Tat Lee, Yuanzhi Li, Weishung Liu, Caio C.~T. Mendes, Anh Nguyen, Eric Price, Gustavo de~Rosa, Olli Saarikivi, and 8 others. 2024.
\newblock \href {https://doi.org/10.48550/ARXIV.2412.08905} {Phi-4 technical report}.
\newblock \emph{CoRR}, abs/2412.08905.

\bibitem[{Akkus et~al.(2025)Akkus, Li, Chu, Backes, Zhang, and Sav}]{DBLP:journals/corr/abs-2409-11423}
Atilla Akkus, Mingjie Li, Junjie Chu, Michael Backes, Yang Zhang, and Sinem Sav. 2025.
\newblock \href {https://doi.org/10.48550/ARXIV.2409.11423} {Generated data with fake privacy: Hidden dangers of fine-tuning large language models on generated data}.
\newblock \emph{CoRR}, abs/2409.11423.

\bibitem[{Akter et~al.(2024)Akter, Prabhumoye, Kamalu, Satheesh, Nyberg, Patwary, Shoeybi, and Catanzaro}]{DBLP:journals/corr/abs-2410-12881}
Syeda~Nahida Akter, Shrimai Prabhumoye, John Kamalu, Sanjeev Satheesh, Eric Nyberg, Mostofa Patwary, Mohammad Shoeybi, and Bryan Catanzaro. 2024.
\newblock \href {https://doi.org/10.48550/ARXIV.2410.12881} {{MIND:} math informed synthetic dialogues for pretraining llms}.
\newblock \emph{CoRR}, abs/2410.12881.

\bibitem[{Allal et~al.(2023)Allal, Li, Kocetkov, Mou, Akiki, Ferrandis, Muennighoff, Mishra, Gu, Dey, Umapathi, Anderson, Zi, Lamy{-}Poirier, Schoelkopf, Troshin, Abulkhanov, Romero, Lappert, Toni, del R{\'{\i}}o, Liu, Bose, Bhattacharyya, Zhuo, Yu, Villegas, Zocca, Mangrulkar, Lansky, Nguyen, Contractor, Villa, Li, Bahdanau, Jernite, Hughes, Fried, Guha, de~Vries, and von Werra}]{DBLP:journals/corr/abs-2301-03988}
Loubna~Ben Allal, Raymond Li, Denis Kocetkov, Chenghao Mou, Christopher Akiki, Carlos~Mu{\~{n}}oz Ferrandis, Niklas Muennighoff, Mayank Mishra, Alex Gu, Manan Dey, Logesh~Kumar Umapathi, Carolyn~Jane Anderson, Yangtian Zi, Joel Lamy{-}Poirier, Hailey Schoelkopf, Sergey Troshin, Dmitry Abulkhanov, Manuel Romero, Michael Lappert, and 22 others. 2023.
\newblock \href {https://doi.org/10.48550/ARXIV.2301.03988} {Santacoder: don't reach for the stars!}
\newblock \emph{CoRR}, abs/2301.03988.

\bibitem[{Anthropic(2025)}]{anthropic2025claude37}
Anthropic. 2025.
\newblock Claude 3.7 sonnet: Hybrid reasoning model.
\newblock \url{https://www.anthropic.com/news/introducing-citations-api}.

\bibitem[{Ashuri and Halperin(2024)}]{ashuri2024online}
Tamar Ashuri and Ruth Halperin. 2024.
\newblock Online self-disclosure: An interdisciplinary literature review of 10 years of research.
\newblock \emph{New Media \& Society}, page 14614448241247313.

\bibitem[{Bai et~al.(2024)Bai, Zhang, Lv, Zheng, Zhu, Hou, Dong, Tang, and Li}]{DBLP:journals/corr/abs-2408-07055}
Yushi Bai, Jiajie Zhang, Xin Lv, Linzhi Zheng, Siqi Zhu, Lei Hou, Yuxiao Dong, Jie Tang, and Juanzi Li. 2024.
\newblock \href {https://doi.org/10.48550/ARXIV.2408.07055} {Longwriter: Unleashing 10,000+ word generation from long context llms}.
\newblock \emph{CoRR}, abs/2408.07055.

\bibitem[{Bassily et~al.(2014)Bassily, Smith, and Thakurta}]{DBLP:conf/focs/BassilyST14}
Raef Bassily, Adam~D. Smith, and Abhradeep Thakurta. 2014.
\newblock \href {https://doi.org/10.1109/FOCS.2014.56} {Private empirical risk minimization: Efficient algorithms and tight error bounds}.
\newblock In \emph{55th {IEEE} Annual Symposium on Foundations of Computer Science, {FOCS} 2014, Philadelphia, PA, USA, October 18-21, 2014}, pages 464--473. {IEEE} Computer Society.

\bibitem[{Bommasani et~al.(2019)Bommasani, Wu, and Schofield}]{bommasani2019towards}
Rishi Bommasani, Steven Wu, and Xanda Schofield. 2019.
\newblock Towards private synthetic text generation.
\newblock In \emph{NeurIPS 2019 Machine Learning with Guarantees Workshop}.

\bibitem[{Cao et~al.(2023)Cao, Li, Liu, Yan, Dai, Yu, and Sun}]{DBLP:journals/corr/abs-2303-04226}
Yihan Cao, Siyu Li, Yixin Liu, Zhiling Yan, Yutong Dai, Philip~S. Yu, and Lichao Sun. 2023.
\newblock \href {https://doi.org/10.48550/ARXIV.2303.04226} {A comprehensive survey of ai-generated content {(AIGC):} {A} history of generative {AI} from {GAN} to chatgpt}.
\newblock \emph{CoRR}, abs/2303.04226.

\bibitem[{Carlini et~al.(2022)Carlini, Chien, Nasr, Song, Terzis, and Tram{\`{e}}r}]{DBLP:conf/sp/CarliniCN0TT22}
Nicholas Carlini, Steve Chien, Milad Nasr, Shuang Song, Andreas Terzis, and Florian Tram{\`{e}}r. 2022.
\newblock \href {https://doi.org/10.1109/SP46214.2022.9833649} {Membership inference attacks from first principles}.
\newblock In \emph{43rd {IEEE} Symposium on Security and Privacy, {SP} 2022, San Francisco, CA, USA, May 22-26, 2022}, pages 1897--1914. {IEEE}.

\bibitem[{Carlini et~al.(2021)Carlini, Tram{\`{e}}r, Wallace, Jagielski, Herbert{-}Voss, Lee, Roberts, Brown, Song, Erlingsson, Oprea, and Raffel}]{DBLP:conf/uss/CarliniTWJHLRBS21}
Nicholas Carlini, Florian Tram{\`{e}}r, Eric Wallace, Matthew Jagielski, Ariel Herbert{-}Voss, Katherine Lee, Adam Roberts, Tom~B. Brown, Dawn Song, {\'{U}}lfar Erlingsson, Alina Oprea, and Colin Raffel. 2021.
\newblock \href {https://www.usenix.org/conference/usenixsecurity21/presentation/carlini-extracting} {Extracting training data from large language models}.
\newblock In \emph{30th {USENIX} Security Symposium, {USENIX} Security 2021, August 11-13, 2021}, pages 2633--2650. {USENIX} Association.

\bibitem[{Choudhury and De(2014)}]{DBLP:conf/icwsm/ChoudhuryD14}
Munmun~De Choudhury and Sushovan De. 2014.
\newblock \href {http://www.aaai.org/ocs/index.php/ICWSM/ICWSM14/paper/view/8075} {Mental health discourse on reddit: Self-disclosure, social support, and anonymity}.
\newblock In \emph{Proceedings of the Eighth International Conference on Weblogs and Social Media, {ICWSM} 2014, Ann Arbor, Michigan, USA, June 1-4, 2014}. The {AAAI} Press.

\bibitem[{DeepSeek{-}AI et~al.(2025)DeepSeek{-}AI, Guo, Yang, Zhang, Song, Zhang, Xu, Zhu, Ma, Wang, Bi, Zhang, Yu, Wu, Wu, Gou, Shao, Li, Gao, Liu, Xue, Wang, Wu, Feng, Lu, Zhao, Deng, Zhang, Ruan, Dai, Chen, Ji, Li, Lin, Dai, Luo, Hao, Chen, Li, Zhang, Bao, Xu, Wang, Ding, Xin, Gao, Qu, Li, Guo, Li, Wang, Chen, Yuan, Qiu, Li, Cai, Ni, Liang, Chen, Dong, Hu, Gao, Guan, Huang, Yu, Wang, Zhang, Zhao, Wang, Zhang, Xu, Xia, Zhang, Zhang, Tang, Li, Wang, Li, Tian, Huang, Zhang, Wang, Chen, Du, Ge, Zhang, Pan, Wang, Chen, Jin, Chen, Lu, Zhou, Chen, Ye, Wang, Yu, Zhou, Pan, and Li}]{DBLP:journals/corr/abs-2501-12948}
DeepSeek{-}AI, Daya Guo, Dejian Yang, Haowei Zhang, Junxiao Song, Ruoyu Zhang, Runxin Xu, Qihao Zhu, Shirong Ma, Peiyi Wang, Xiao Bi, Xiaokang Zhang, Xingkai Yu, Yu~Wu, Z.~F. Wu, Zhibin Gou, Zhihong Shao, Zhuoshu Li, Ziyi Gao, and 81 others. 2025.
\newblock \href {https://doi.org/10.48550/ARXIV.2501.12948} {Deepseek-r1: Incentivizing reasoning capability in llms via reinforcement learning}.
\newblock \emph{CoRR}, abs/2501.12948.

\bibitem[{DeSalvo et~al.(2024)DeSalvo, Kagy, Karydas, Rostamizadeh, and Kumar}]{DBLP:journals/corr/abs-2410-16534}
Giulia DeSalvo, Jean{-}Fran{\c{c}}ois Kagy, Lazaros Karydas, Afshin Rostamizadeh, and Sanjiv Kumar. 2024.
\newblock \href {https://doi.org/10.48550/ARXIV.2410.16534} {No more hard prompts: Softsrv prompting for synthetic data generation}.
\newblock \emph{CoRR}, abs/2410.16534.

\bibitem[{Dou et~al.(2024)Dou, Krsek, Naous, Kabra, Das, Ritter, and Xu}]{DBLP:conf/acl/DouKNKDRX24}
Yao Dou, Isadora Krsek, Tarek Naous, Anubha Kabra, Sauvik Das, Alan Ritter, and Wei Xu. 2024.
\newblock \href {https://doi.org/10.18653/V1/2024.ACL-LONG.741} {Reducing privacy risks in online self-disclosures with language models}.
\newblock In \emph{Proceedings of the 62nd Annual Meeting of the Association for Computational Linguistics (Volume 1: Long Papers), {ACL} 2024, Bangkok, Thailand, August 11-16, 2024}, pages 13732--13754. Association for Computational Linguistics.

\bibitem[{Du et~al.(2024)Du, Kim, Squicciarini, and Rajtmajer}]{du2024toward}
Tingting Du, Jiyoon Kim, Anna Squicciarini, and Sarah Rajtmajer. 2024.
\newblock Toward context-aware privacy enhancing technologies for online self-disclosure.
\newblock In \emph{Proceedings of the AAAI Conference on Human Computation and Crowdsourcing}, volume~12, pages 44--54.

\bibitem[{Duan et~al.(2024)Duan, Suri, Mireshghallah, Min, Shi, Zettlemoyer, Tsvetkov, Choi, Evans, and Hajishirzi}]{DBLP:journals/corr/abs-2402-07841}
Michael Duan, Anshuman Suri, Niloofar Mireshghallah, Sewon Min, Weijia Shi, Luke Zettlemoyer, Yulia Tsvetkov, Yejin Choi, David Evans, and Hannaneh Hajishirzi. 2024.
\newblock \href {https://doi.org/10.48550/ARXIV.2402.07841} {Do membership inference attacks work on large language models?}
\newblock \emph{CoRR}, abs/2402.07841.

\bibitem[{Dwork and Roth(2014)}]{DBLP:journals/fttcs/DworkR14}
Cynthia Dwork and Aaron Roth. 2014.
\newblock \href {https://doi.org/10.1561/0400000042} {The algorithmic foundations of differential privacy}.
\newblock \emph{Found. Trends Theor. Comput. Sci.}, 9(3-4):211--407.

\bibitem[{Edemacu and Wu(2024)}]{DBLP:journals/corr/abs-2404-06001}
Kennedy Edemacu and Xintao Wu. 2024.
\newblock \href {https://doi.org/10.48550/ARXIV.2404.06001} {Privacy preserving prompt engineering: {A} survey}.
\newblock \emph{CoRR}, abs/2404.06001.

\bibitem[{Frikha et~al.(2024)Frikha, Walha, Nakka, Mendes, Jiang, and Zhou}]{DBLP:journals/corr/abs-2407-02956}
Ahmed Frikha, Nassim Walha, Krishna~Kanth Nakka, Ricardo Mendes, Xue Jiang, and Xuebing Zhou. 2024.
\newblock \href {https://doi.org/10.48550/ARXIV.2407.02956} {Incognitext: Privacy-enhancing conditional text anonymization via llm-based private attribute randomization}.
\newblock \emph{CoRR}, abs/2407.02956.

\bibitem[{Gruzd and Hern{\'{a}}ndez{-}Garc{\'{\i}}a(2018)}]{DBLP:journals/cbsn/GruzdH18}
Anatoliy~A. Gruzd and {\'{A}}ngel Hern{\'{a}}ndez{-}Garc{\'{\i}}a. 2018.
\newblock \href {https://doi.org/10.1089/CYBER.2017.0709} {Privacy concerns and self-disclosure in private and public uses of social media}.
\newblock \emph{Cyberpsychology Behav. Soc. Netw.}, 21(7):418--428.

\bibitem[{Gu et~al.(2022)Gu, Tinn, Cheng, Lucas, Usuyama, Liu, Naumann, Gao, and Poon}]{DBLP:journals/health/GuTCLULNGP22}
Yu~Gu, Robert Tinn, Hao Cheng, Michael Lucas, Naoto Usuyama, Xiaodong Liu, Tristan Naumann, Jianfeng Gao, and Hoifung Poon. 2022.
\newblock \href {https://doi.org/10.1145/3458754} {Domain-specific language model pretraining for biomedical natural language processing}.
\newblock \emph{{ACM} Trans. Comput. Heal.}, 3(1):2:1--2:23.

\bibitem[{H{\"{a}}m{\"{a}}l{\"{a}}inen et~al.(2023)H{\"{a}}m{\"{a}}l{\"{a}}inen, Tavast, and Kunnari}]{DBLP:conf/chi/HamalainenTK23}
Perttu H{\"{a}}m{\"{a}}l{\"{a}}inen, Mikke Tavast, and Anton Kunnari. 2023.
\newblock \href {https://doi.org/10.1145/3544548.3580688} {Evaluating large language models in generating synthetic {HCI} research data: a case study}.
\newblock In \emph{Proceedings of the 2023 {CHI} Conference on Human Factors in Computing Systems, {CHI} 2023, Hamburg, Germany, April 23-28, 2023}, pages 433:1--433:19. {ACM}.

\bibitem[{Hartmann et~al.(2023)Hartmann, Heitmann, Siebert, and Schamp}]{hartmann2023}
Jochen Hartmann, Mark Heitmann, Christian Siebert, and Christina Schamp. 2023.
\newblock \href {https://doi.org/10.1016/j.ijresmar.2022.05.005} {More than a feeling: Accuracy and application of sentiment analysis}.
\newblock \emph{International Journal of Research in Marketing}, 40(1):75--87.

\bibitem[{He et~al.(2023)He, Zhang, Wang, Kumar, Cho, Glass, and Tsvetkov}]{DBLP:conf/acl/HeZ0KCGT23}
Tianxing He, Jingyu Zhang, Tianle Wang, Sachin Kumar, Kyunghyun Cho, James~R. Glass, and Yulia Tsvetkov. 2023.
\newblock \href {https://doi.org/10.18653/V1/2023.ACL-LONG.674} {On the blind spots of model-based evaluation metrics for text generation}.
\newblock In \emph{Proceedings of the 61st Annual Meeting of the Association for Computational Linguistics (Volume 1: Long Papers), {ACL} 2023, Toronto, Canada, July 9-14, 2023}, pages 12067--12097. Association for Computational Linguistics.

\bibitem[{Holtzman et~al.(2020)Holtzman, Buys, Du, Forbes, and Choi}]{DBLP:conf/iclr/HoltzmanBDFC20}
Ari Holtzman, Jan Buys, Li~Du, Maxwell Forbes, and Yejin Choi. 2020.
\newblock \href {https://openreview.net/forum?id=rygGQyrFvH} {The curious case of neural text degeneration}.
\newblock In \emph{8th International Conference on Learning Representations, {ICLR} 2020, Addis Ababa, Ethiopia, April 26-30, 2020}. OpenReview.net.

\bibitem[{Hou et~al.(2024)Hou, Shrivastava, Zhan, Conway, Le, Sagar, Fanti, and Lazar}]{DBLP:conf/icml/HouSZCLSFL24}
Charlie Hou, Akshat Shrivastava, Hongyuan Zhan, Rylan Conway, Trang Le, Adithya Sagar, Giulia Fanti, and Daniel Lazar. 2024.
\newblock \href {https://openreview.net/forum?id=3WCvnkHnxV} {Pre-text: Training language models on private federated data in the age of llms}.
\newblock In \emph{Forty-first International Conference on Machine Learning, {ICML} 2024, Vienna, Austria, July 21-27, 2024}. OpenReview.net.

\bibitem[{Kuo et~al.(2024)Kuo, Gallego, and Jorm}]{DBLP:journals/corr/abs-2410-16811}
Nicholas~I{-}Hsien Kuo, Blanca Gallego, and Louisa Jorm. 2024.
\newblock \href {https://doi.org/10.48550/ARXIV.2410.16811} {Masked clinical modelling: {A} framework for synthetic and augmented survival data generation}.
\newblock \emph{CoRR}, abs/2410.16811.

\bibitem[{Kurakin et~al.(2023{\natexlab{a}})Kurakin, Ponomareva, Syed, MacDermed, and Terzis}]{kurakin2023harnessing}
Alexey Kurakin, Natalia Ponomareva, Umar Syed, Liam MacDermed, and Andreas Terzis. 2023{\natexlab{a}}.
\newblock Harnessing large-language models to generate private synthetic text.
\newblock \emph{arXiv preprint arXiv:2306.01684}.

\bibitem[{Kurakin et~al.(2023{\natexlab{b}})Kurakin, Ponomareva, Syed, MacDermed, and Terzis}]{DBLP:journals/corr/abs-2306-01684}
Alexey Kurakin, Natalia Ponomareva, Umar Syed, Liam MacDermed, and Andreas Terzis. 2023{\natexlab{b}}.
\newblock \href {https://doi.org/10.48550/ARXIV.2306.01684} {Harnessing large-language models to generate private synthetic text}.
\newblock \emph{CoRR}, abs/2306.01684.

\bibitem[{Lewis et~al.(2019)Lewis, Liu, Goyal, Ghazvininejad, Mohamed, Levy, Stoyanov, and Zettlemoyer}]{DBLP:journals/corr/abs-1910-13461}
Mike Lewis, Yinhan Liu, Naman Goyal, Marjan Ghazvininejad, Abdelrahman Mohamed, Omer Levy, Veselin Stoyanov, and Luke Zettlemoyer. 2019.
\newblock \href {https://arxiv.org/abs/1910.13461} {{BART:} denoising sequence-to-sequence pre-training for natural language generation, translation, and comprehension}.
\newblock \emph{CoRR}, abs/1910.13461.

\bibitem[{Li et~al.(2024{\natexlab{a}})Li, Hong, Xie, Tan, Xin, Hou, Yin, Wang, Hendrycks, Wang, Li, He, and Song}]{DBLP:journals/pvldb/LiHXTXHYWHWLHS24}
Qinbin Li, Junyuan Hong, Chulin Xie, Jeffrey Tan, Rachel Xin, Junyi Hou, Xavier Yin, Zhun Wang, Dan Hendrycks, Zhangyang Wang, Bo~Li, Bingsheng He, and Dawn Song. 2024{\natexlab{a}}.
\newblock \href {https://doi.org/10.14778/3681954.3681994} {{LLM-PBE:} assessing data privacy in large language models}.
\newblock \emph{Proc. {VLDB} Endow.}, 17(11):3201--3214.

\bibitem[{Li et~al.(2024{\natexlab{b}})Li, Yu, and Xiong}]{DBLP:journals/corr/abs-2410-14208}
Xiaochuan Li, Zichun Yu, and Chenyan Xiong. 2024{\natexlab{b}}.
\newblock \href {https://doi.org/10.48550/ARXIV.2410.14208} {Montessori-instruct: Generate influential training data tailored for student learning}.
\newblock \emph{CoRR}, abs/2410.14208.

\bibitem[{Long et~al.(2024)Long, Wang, Xiao, Zhao, Ding, Chen, and Wang}]{DBLP:conf/acl/LongWXZDCW24}
Lin Long, Rui Wang, Ruixuan Xiao, Junbo Zhao, Xiao Ding, Gang Chen, and Haobo Wang. 2024.
\newblock \href {https://doi.org/10.18653/V1/2024.FINDINGS-ACL.658} {On llms-driven synthetic data generation, curation, and evaluation: {A} survey}.
\newblock In \emph{Findings of the Association for Computational Linguistics, {ACL} 2024, Bangkok, Thailand and virtual meeting, August 11-16, 2024}, pages 11065--11082. Association for Computational Linguistics.

\bibitem[{Lu et~al.(2023)Lu, Wang, and Wei}]{DBLP:journals/corr/abs-2302-04062}
Yingzhou Lu, Huazheng Wang, and Wenqi Wei. 2023.
\newblock \href {https://doi.org/10.48550/ARXIV.2302.04062} {Machine learning for synthetic data generation: a review}.
\newblock \emph{CoRR}, abs/2302.04062.

\bibitem[{Malladi et~al.(2023)Malladi, Gao, Nichani, Damian, Lee, Chen, and Arora}]{DBLP:conf/nips/MalladiGNDL0A23}
Sadhika Malladi, Tianyu Gao, Eshaan Nichani, Alex Damian, Jason~D. Lee, Danqi Chen, and Sanjeev Arora. 2023.
\newblock \href {http://papers.nips.cc/paper\_files/paper/2023/hash/a627810151be4d13f907ac898ff7e948-Abstract-Conference.html} {Fine-tuning language models with just forward passes}.
\newblock In \emph{Advances in Neural Information Processing Systems 36: Annual Conference on Neural Information Processing Systems 2023, NeurIPS 2023, New Orleans, LA, USA, December 10 - 16, 2023}.

\bibitem[{Mattern et~al.(2022)Mattern, Jin, Weggenmann, Sch{\"{o}}lkopf, and Sachan}]{DBLP:conf/emnlp/MatternJWSS22}
Justus Mattern, Zhijing Jin, Benjamin Weggenmann, Bernhard Sch{\"{o}}lkopf, and Mrinmaya Sachan. 2022.
\newblock \href {https://doi.org/10.18653/V1/2022.EMNLP-MAIN.323} {Differentially private language models for secure data sharing}.
\newblock In \emph{Proceedings of the 2022 Conference on Empirical Methods in Natural Language Processing, {EMNLP} 2022, Abu Dhabi, United Arab Emirates, December 7-11, 2022}, pages 4860--4873. Association for Computational Linguistics.

\bibitem[{Mattern et~al.(2023)Mattern, Mireshghallah, Jin, Sch{\"{o}}lkopf, Sachan, and Berg{-}Kirkpatrick}]{DBLP:conf/acl/MatternMJSSB23}
Justus Mattern, Fatemehsadat Mireshghallah, Zhijing Jin, Bernhard Sch{\"{o}}lkopf, Mrinmaya Sachan, and Taylor Berg{-}Kirkpatrick. 2023.
\newblock \href {https://doi.org/10.18653/V1/2023.FINDINGS-ACL.719} {Membership inference attacks against language models via neighbourhood comparison}.
\newblock In \emph{Findings of the Association for Computational Linguistics: {ACL} 2023, Toronto, Canada, July 9-14, 2023}, pages 11330--11343. Association for Computational Linguistics.

\bibitem[{Meta(2024)}]{meta2024llama33}
Meta. 2024.
\newblock Llama 3.3: Multilingual large language model.
\newblock \url{https://huggingface.co/meta-llama/Llama-3.3-70B-Instruct}.

\bibitem[{Mireshghallah et~al.(2022)Mireshghallah, Uniyal, Wang, Evans, and Berg{-}Kirkpatrick}]{DBLP:journals/corr/abs-2205-12506}
Fatemehsadat Mireshghallah, Archit Uniyal, Tianhao Wang, David Evans, and Taylor Berg{-}Kirkpatrick. 2022.
\newblock \href {https://doi.org/10.48550/ARXIV.2205.12506} {Memorization in {NLP} fine-tuning methods}.
\newblock \emph{CoRR}, abs/2205.12506.

\bibitem[{Mironov(2017)}]{DBLP:conf/csfw/Mironov17}
Ilya Mironov. 2017.
\newblock \href {https://doi.org/10.1109/CSF.2017.11} {R{\'{e}}nyi differential privacy}.
\newblock In \emph{30th {IEEE} Computer Security Foundations Symposium, {CSF} 2017, Santa Barbara, CA, USA, August 21-25, 2017}, pages 263--275. {IEEE} Computer Society.

\bibitem[{Montahaei et~al.(2019)Montahaei, Alihosseini, and Baghshah}]{DBLP:journals/corr/abs-1904-03971}
Ehsan Montahaei, Danial Alihosseini, and Mahdieh~Soleymani Baghshah. 2019.
\newblock \href {https://arxiv.org/abs/1904.03971} {Jointly measuring diversity and quality in text generation models}.
\newblock \emph{CoRR}, abs/1904.03971.

\bibitem[{OpenAI(2023)}]{DBLP:journals/corr/abs-2303-08774}
OpenAI. 2023.
\newblock \href {https://doi.org/10.48550/ARXIV.2303.08774} {{GPT-4} technical report}.
\newblock \emph{CoRR}, abs/2303.08774.

\bibitem[{OpenAI(2025)}]{openai2025gpt5}
OpenAI. 2025.
\newblock Introducing gpt-5.
\newblock \url{https://openai.com/index/introducing-gpt-5/}.

\bibitem[{Papernot et~al.(2017)Papernot, Abadi, Erlingsson, Goodfellow, and Talwar}]{DBLP:conf/iclr/PapernotAEGT17}
Nicolas Papernot, Mart{\'{\i}}n Abadi, {\'{U}}lfar Erlingsson, Ian~J. Goodfellow, and Kunal Talwar. 2017.
\newblock \href {https://openreview.net/forum?id=HkwoSDPgg} {Semi-supervised knowledge transfer for deep learning from private training data}.
\newblock In \emph{5th International Conference on Learning Representations, {ICLR} 2017, Toulon, France, April 24-26, 2017, Conference Track Proceedings}. OpenReview.net.

\bibitem[{Papernot et~al.(2018)Papernot, Song, Mironov, Raghunathan, Talwar, and Erlingsson}]{DBLP:conf/iclr/PapernotSMRTE18}
Nicolas Papernot, Shuang Song, Ilya Mironov, Ananth Raghunathan, Kunal Talwar, and {\'{U}}lfar Erlingsson. 2018.
\newblock \href {https://openreview.net/forum?id=rkZB1XbRZ} {Scalable private learning with {PATE}}.
\newblock In \emph{6th International Conference on Learning Representations, {ICLR} 2018, Vancouver, BC, Canada, April 30 - May 3, 2018, Conference Track Proceedings}. OpenReview.net.

\bibitem[{Reimers and Gurevych(2019)}]{DBLP:conf/emnlp/ReimersG19}
Nils Reimers and Iryna Gurevych. 2019.
\newblock \href {https://doi.org/10.18653/V1/D19-1410} {Sentence-bert: Sentence embeddings using siamese bert-networks}.
\newblock In \emph{Proceedings of the 2019 Conference on Empirical Methods in Natural Language Processing and the 9th International Joint Conference on Natural Language Processing, {EMNLP-IJCNLP} 2019, Hong Kong, China, November 3-7, 2019}, pages 3980--3990. Association for Computational Linguistics.

\bibitem[{Rumshisky et~al.(2016)Rumshisky, Ghassemi, Naumann, Szolovits, Castro, McCoy, and Perlis}]{rumshisky2016predicting}
Anna Rumshisky, Marzyeh Ghassemi, Tristan Naumann, Peter Szolovits, VM~Castro, TH~McCoy, and RH~Perlis. 2016.
\newblock Predicting early psychiatric readmission with natural language processing of narrative discharge summaries.
\newblock \emph{Translational psychiatry}, 6(10):e921--e921.

\bibitem[{Sanh et~al.(2019)Sanh, Debut, Chaumond, and Wolf}]{sanh2019distilbert}
Victor Sanh, Lysandre Debut, Julien Chaumond, and Thomas Wolf. 2019.
\newblock Distilbert, a distilled version of bert: smaller, faster, cheaper and lighter.
\newblock In \emph{NeurIPS EMC2 Workshop}.

\bibitem[{Shirgaonkar et~al.(2024)Shirgaonkar, Pandey, Abay, Aktas, and Aski}]{DBLP:journals/corr/abs-2410-18588}
Anup Shirgaonkar, Nikhil Pandey, Nazmiye~Ceren Abay, Tolga Aktas, and Vijay Aski. 2024.
\newblock \href {https://doi.org/10.48550/ARXIV.2410.18588} {Knowledge distillation using frontier open-source llms: Generalizability and the role of synthetic data}.
\newblock \emph{CoRR}, abs/2410.18588.

\bibitem[{Shokri et~al.(2017)Shokri, Stronati, Song, and Shmatikov}]{DBLP:conf/sp/ShokriSSS17}
Reza Shokri, Marco Stronati, Congzheng Song, and Vitaly Shmatikov. 2017.
\newblock \href {https://doi.org/10.1109/SP.2017.41} {Membership inference attacks against machine learning models}.
\newblock In \emph{2017 {IEEE} Symposium on Security and Privacy, {SP} 2017, San Jose, CA, USA, May 22-26, 2017}, pages 3--18. {IEEE} Computer Society.

\bibitem[{Tang et~al.(2023)Tang, Han, Jiang, and Hu}]{DBLP:journals/corr/abs-2303-04360}
Ruixiang Tang, Xiaotian Han, Xiaoqian Jiang, and Xia Hu. 2023.
\newblock \href {https://doi.org/10.48550/ARXIV.2303.04360} {Does synthetic data generation of llms help clinical text mining?}
\newblock \emph{CoRR}, abs/2303.04360.

\bibitem[{Turc et~al.(2019)Turc, Chang, Lee, and Toutanova}]{DBLP:journals/corr/abs-1908-08962}
Iulia Turc, Ming{-}Wei Chang, Kenton Lee, and Kristina Toutanova. 2019.
\newblock \href {https://arxiv.org/abs/1908.08962} {Well-read students learn better: The impact of student initialization on knowledge distillation}.
\newblock \emph{CoRR}, abs/1908.08962.

\bibitem[{Walonoski et~al.(2018)Walonoski, Kramer, Nichols, Quina, Moesel, Hall, Duffett, Dube, Gallagher, and McLachlan}]{DBLP:journals/jamia/WalonoskiKNQMHD18}
Jason~A. Walonoski, Mark Kramer, Joseph Nichols, Andre Quina, Chris Moesel, Dylan Hall, Carlton Duffett, Kudakwashe Dube, Thomas Gallagher, and Scott McLachlan. 2018.
\newblock \href {https://doi.org/10.1093/JAMIA/OCX079} {Synthea: An approach, method, and software mechanism for generating synthetic patients and the synthetic electronic health care record}.
\newblock \emph{J. Am. Medical Informatics Assoc.}, 25(3):230--238.

\bibitem[{Wang et~al.(2023)Wang, Chen, Pei, Xie, Kang, Zhang, Xu, Xiong, Dutta, Schaeffer, Truong, Arora, Mazeika, Hendrycks, Lin, Cheng, Koyejo, Song, and Li}]{DBLP:conf/nips/WangCPXKZXXDSTA23}
Boxin Wang, Weixin Chen, Hengzhi Pei, Chulin Xie, Mintong Kang, Chenhui Zhang, Chejian Xu, Zidi Xiong, Ritik Dutta, Rylan Schaeffer, Sang~T. Truong, Simran Arora, Mantas Mazeika, Dan Hendrycks, Zinan Lin, Yu~Cheng, Sanmi Koyejo, Dawn Song, and Bo~Li. 2023.
\newblock \href {http://papers.nips.cc/paper\_files/paper/2023/hash/63cb9921eecf51bfad27a99b2c53dd6d-Abstract-Datasets\_and\_Benchmarks.html} {Decodingtrust: {A} comprehensive assessment of trustworthiness in {GPT} models}.
\newblock In \emph{Advances in Neural Information Processing Systems 36: Annual Conference on Neural Information Processing Systems 2023, NeurIPS 2023, New Orleans, LA, USA, December 10 - 16, 2023}.

\bibitem[{Wen et~al.(2024)Wen, Marchyok, Hong, Geiping, Goldstein, and Carlini}]{DBLP:journals/corr/abs-2404-01231}
Yuxin Wen, Leo Marchyok, Sanghyun Hong, Jonas Geiping, Tom Goldstein, and Nicholas Carlini. 2024.
\newblock \href {https://doi.org/10.48550/ARXIV.2404.01231} {Privacy backdoors: Enhancing membership inference through poisoning pre-trained models}.
\newblock \emph{CoRR}, abs/2404.01231.

\bibitem[{Wolf et~al.(2020)Wolf, Debut, Sanh, Chaumond, Delangue, Moi, Cistac, Rault, Louf, Funtowicz, Davison, Shleifer, von Platen, Ma, Jernite, Plu, Xu, Scao, Gugger, Drame, Lhoest, and Rush}]{DBLP:conf/emnlp/WolfDSCDMCRLFDS20}
Thomas Wolf, Lysandre Debut, Victor Sanh, Julien Chaumond, Clement Delangue, Anthony Moi, Pierric Cistac, Tim Rault, R{\'{e}}mi Louf, Morgan Funtowicz, Joe Davison, Sam Shleifer, Patrick von Platen, Clara Ma, Yacine Jernite, Julien Plu, Canwen Xu, Teven~Le Scao, Sylvain Gugger, and 3 others. 2020.
\newblock \href {https://doi.org/10.18653/V1/2020.EMNLP-DEMOS.6} {Transformers: State-of-the-art natural language processing}.
\newblock In \emph{Proceedings of the 2020 Conference on Empirical Methods in Natural Language Processing: System Demonstrations, {EMNLP} 2020 - Demos, Online, November 16-20, 2020}, pages 38--45. Association for Computational Linguistics.

\bibitem[{Xie et~al.(2024)Xie, Lin, Backurs, Gopi, Yu, Inan, Nori, Jiang, Zhang, Lee, Li, and Yekhanin}]{DBLP:conf/icml/Xie0BGYINJZL0Y24}
Chulin Xie, Zinan Lin, Arturs Backurs, Sivakanth Gopi, Da~Yu, Huseyin~A. Inan, Harsha Nori, Haotian Jiang, Huishuai Zhang, Yin~Tat Lee, Bo~Li, and Sergey Yekhanin. 2024.
\newblock \href {https://openreview.net/forum?id=LWD7upg1ob} {Differentially private synthetic data via foundation model apis 2: Text}.
\newblock In \emph{Forty-first International Conference on Machine Learning, {ICML} 2024, Vienna, Austria, July 21-27, 2024}. OpenReview.net.

\bibitem[{Yang et~al.(2025)Yang, Zhu, and Gurevych}]{DBLP:conf/acl/0004ZG25}
Tianyu Yang, Xiaodan Zhu, and Iryna Gurevych. 2025.
\newblock \href {https://aclanthology.org/2025.acl-long.1404/} {Robust utility-preserving text anonymization based on large language models}.
\newblock In \emph{Proceedings of the 63rd Annual Meeting of the Association for Computational Linguistics (Volume 1: Long Papers), {ACL} 2025, Vienna, Austria, July 27 - August 1, 2025}, pages 28922--28941. Association for Computational Linguistics.

\bibitem[{Yu et~al.(2023)Yu, Backurs, Gopi, Inan, Kulkarni, Lin, Xie, Zhang, and Zhang}]{yu2023pubmed}
Da~Yu, Arturs Backurs, Sivakanth Gopi, Huseyin Inan, Janardhan Kulkarni, Zinan Lin, Chulin Xie, Huishuai Zhang, and Wanrong Zhang. 2023.
\newblock Training private and efficient language models with synthetic data from llms.
\newblock In \emph{Socially Responsible Language Modelling Research}.

\bibitem[{Yu et~al.(2022)Yu, Naik, Backurs, Gopi, Inan, Kamath, Kulkarni, Lee, Manoel, Wutschitz, Yekhanin, and Zhang}]{DBLP:conf/iclr/YuNBGI0KLMWYZ22}
Da~Yu, Saurabh Naik, Arturs Backurs, Sivakanth Gopi, Huseyin~A. Inan, Gautam Kamath, Janardhan Kulkarni, Yin~Tat Lee, Andre Manoel, Lukas Wutschitz, Sergey Yekhanin, and Huishuai Zhang. 2022.
\newblock \href {https://openreview.net/forum?id=Q42f0dfjECO} {Differentially private fine-tuning of language models}.
\newblock In \emph{The Tenth International Conference on Learning Representations, {ICLR} 2022, Virtual Event, April 25-29, 2022}. OpenReview.net.

\bibitem[{Yue et~al.(2023)Yue, Inan, Li, Kumar, McAnallen, Shajari, Sun, Levitan, and Sim}]{DBLP:conf/acl/YueILKMS0LS23}
Xiang Yue, Huseyin~A. Inan, Xuechen Li, Girish Kumar, Julia McAnallen, Hoda Shajari, Huan Sun, David Levitan, and Robert Sim. 2023.
\newblock \href {https://doi.org/10.18653/V1/2023.ACL-LONG.74} {Synthetic text generation with differential privacy: {A} simple and practical recipe}.
\newblock In \emph{Proceedings of the 61st Annual Meeting of the Association for Computational Linguistics (Volume 1: Long Papers), {ACL} 2023, Toronto, Canada, July 9-14, 2023}, pages 1321--1342. Association for Computational Linguistics.

\bibitem[{Yukhymenko et~al.(2024)Yukhymenko, Staab, Vero, and Vechev}]{DBLP:journals/corr/abs-2406-07217}
Hanna Yukhymenko, Robin Staab, Mark Vero, and Martin~T. Vechev. 2024.
\newblock \href {https://doi.org/10.48550/ARXIV.2406.07217} {A synthetic dataset for personal attribute inference}.
\newblock \emph{CoRR}, abs/2406.07217.

\bibitem[{Zhu et~al.(2018)Zhu, Lu, Zheng, Guo, Zhang, Wang, and Yu}]{DBLP:conf/sigir/ZhuLZGZWY18}
Yaoming Zhu, Sidi Lu, Lei Zheng, Jiaxian Guo, Weinan Zhang, Jun Wang, and Yong Yu. 2018.
\newblock \href {https://doi.org/10.1145/3209978.3210080} {Texygen: {A} benchmarking platform for text generation models}.
\newblock In \emph{The 41st International {ACM} {SIGIR} Conference on Research {\&} Development in Information Retrieval, {SIGIR} 2018, Ann Arbor, MI, USA, July 08-12, 2018}, pages 1097--1100. {ACM}.

\bibitem[{Zhu et~al.(2020)Zhu, Yu, Chandraker, and Wang}]{DBLP:conf/cvpr/00050CW20}
Yuqing Zhu, Xiang Yu, Manmohan Chandraker, and Yu{-}Xiang Wang. 2020.
\newblock \href {https://doi.org/10.1109/CVPR42600.2020.01187} {Private-knn: Practical differential privacy for computer vision}.
\newblock In \emph{2020 {IEEE/CVF} Conference on Computer Vision and Pattern Recognition, {CVPR} 2020, Seattle, WA, USA, June 13-19, 2020}, pages 11851--11859. Computer Vision Foundation / {IEEE}.

\end{thebibliography}

\clearpage
\appendix

\section{Preliminaries}
\label{appendix-preliminaries} 

\subsection{\added{Differential Privacy}}
\added{Differential privacy (DP) \cite{DBLP:journals/fttcs/DworkR14,rumshisky2016predicting} offers strong guarantees for protecting individual privacy in the context of data analysis and machine learning. The formal definition of DP is given by:}

\added{For all data sets $\mathcal{D}$ and $\mathcal{D^{\prime}}$ differing in one element, and for all subsets \(S \subseteq {\textrm{Range}}(f)\),}
\begingroup
\added{
\[
{Pr}[M(\mathcal{D}) \in S] \leq \exp(\epsilon) \times {Pr}[M(\mathcal{D^{\prime}}) \in S] + \delta
\]
}
\endgroup
\added{where \(M\) is the randomized mechanism providing privacy; \(\epsilon\) and \(\delta\) are privacy parameters, representing the degree of privacy protection.}

\subsection{\added{Gaussian Mechanism}}
\added{The Gaussian mechanism is a commonly used method to achieve DP in the context of noisy data release \cite{DBLP:conf/csfw/Mironov17}. It is defined as follows:}
\begingroup
\added{
\[
M(D) = f(D) + \mathcal{N}(0, \sigma^2I)
\]
}
\endgroup
\added{where \(M(D)\) is the output of the mechanism; \(f(D)\) is the deterministic function applied to the input data \(D\); \(\mathcal{N}(0, \sigma^2I)\) represents Gaussian noise added to the output, with mean \(0\) and covariance matrix \(\sigma^2I\).}

\subsection{Privacy Guarantee of Candidate Selection}
\label{appendix-preliminaries-privacy-guarantee} 
\noindent\textbf{Proposition 1 (DP guarantee of the candidate selection mechanism).}
Let $x$ be a private seed, and let $\mathcal{S}_{\mathrm{abc}}(x)$ be its candidate set with utilities
$u(x,s_j) \in [0,1]$ for $s_j \in \mathcal{S}_{\mathrm{abc}}(x)$.
Assume that $u$ has $\ell_2$-sensitivity $\Delta u = 1$ with respect to neighboring private datasets.
For privacy parameters $\epsilon \in (0,1]$ and $\delta \in (0,1)$, we use the noise scale in Eq.~\ref{eq:gaussian_sigma} and release the DP-selected candidate in Eq.~\ref{eq:dp_selection}. Then the resulting candidate selection mechanism is $(\epsilon,\delta)$-DP with respect to the private dataset.\\

\noindent\textit{Proof.}
The mechanism above is an instantiation of the classical Gaussian mechanism
for a function with sensitivity $\Delta u = 1$. The stated noise scale follows
the standard calibration for $(\epsilon,\delta)$-DP
\cite{DBLP:journals/fttcs/DworkR14}, and is widely used in practice for DP-SGD
\cite{DBLP:conf/ccs/AbadiCGMMT016}. This establishes the $(\epsilon,\delta)$-DP
guarantee for the candidate selection mechanism.

\section{Additional Experimental Details}
\label{appendix-additional-experimental-details}

\subsection{\added{Abstraction Models and Sentiment Alignment}}
\label{appendix-abstraction-alignment}

\added{For each private seed $x$, we use a pretrained abstraction model (facebook/bart-large-cnn~\cite{DBLP:journals/corr/abs-1910-13461}) to generate an oversampled pool of $K\!\ge\!m$ candidates, then prune to $m$ high-quality abstractions $\mathcal{S}_{\mathrm{abc}}(x)=\{s_1,\dots,s_m\}$ that are both semantically faithful and sentiment-consistent with $x$. We compute sentence embeddings with a sentence-transformers encoder (sentence-t5-base~\cite{DBLP:conf/emnlp/ReimersG19}) and measure cosine similarity $\cos(\mathbf e(x),\mathbf e(s))$. A lightweight sentiment model (siebert/sentiment-roberta-large-englishh~\cite{hartmann2023}) provides polarity $y(s)\in\{0,1\}$ and confidence $\mathrm{conf}(s)$. If a seed polarity is available, we prepend a minimal control phrase (\emph{“Keep positive tone:”} or \emph{“Keep negative tone:”}) prior to abstraction.}

\added{Candidates are scored by:}
\begingroup
\added{
\begin{equation}
\small
\begin{split}
\mathrm{score}(s) = & \;\beta \cdot \cos(\mathbf e(x),\mathbf e(s)) \\
& + (1-\beta)\cdot \indic{y(s)=y(x)} \\
& - \lambda \cdot \big(1-\indic{y(s)=y(x)}\big).
\end{split}
\end{equation}
}
\endgroup
\added{with a confidence gate $\mathrm{conf}(s)\ge\kappa$ when enforcing agreement. We first decode with beam search, if no beam candidate satisfies agreement at confidence $\kappa$, we perform a single sampling retry and re-score. The top $m$ candidates by $\mathrm{score}(\cdot)$ form $\mathcal{S}_{\mathrm{abc}}(x)$, which is the exact input to the DP selection step (\S\ref{subsec:dp_selection}). The hyperparameters used for abstraction candidate generation are summarized in Table~\ref{tab:abstraction_params}.}

\begin{table}
    \centering  
    \fontsize{8}{9.5}\selectfont 
    \begin{tabular}{
        >{\centering\arraybackslash}p{2.8cm} |
        >{\centering\arraybackslash}p{2.0cm}}
        \toprule
        \textbf{Parameter} & \textbf{Value} \\
        \midrule       
        $m$ (candidates for DP) & 5 \\
        \midrule
        $K$ (oversampling) & 10 \\
        \midrule
        $\beta$ (sim. weight) & 0.75 \\
        \midrule
        $\lambda$ (flip penalty) & 0.15 \\
        \midrule
        $\kappa$ (min conf) & 0.55 \\
        \midrule
        Decoding attempts & 2 (beam + retry) \\
        \midrule
        Max / Min length & 150 / 50 tokens \\
        \bottomrule
    \end{tabular}
    \caption{\added{Hyperparameters for Sentiment Alignment in Abstraction Phase.}}
    \label{tab:abstraction_params}
\end{table}

\subsection{Prompt Design}
\label{appendix-prompt-design}

Table~\ref{tab:prompt_design} presents the prompt templates used in RPSG. To generate a synthetic variant from an abstracted private candidate, RPSG uses the prompt listed under\textit{ Generating Synthetic Variant}. To generate a synthetic sample from the synthetic variant, it uses the prompt under \textit{Generating Synthetic data}.

\subsection{\added{Privacy Alignment and Calibration}}
\label{appendix-privacy-alignment}
\added{We evaluate DP-SGD, AUG-PE, and RPSG under matched privacy guarantees at 
$\epsilon \in \{4,2,1\}$ with a common $\delta = 1/(N \log N)$, where $N{=}8948$ (the size of our Reddit training dataset). 
Because the three methods employ different mechanisms and privacy accountants, 
the mapping from $(\epsilon,\delta)$ to the corresponding noise multiplier 
$\sigma$ differs. To ensure fairness, we align comparisons on the formal privacy 
guarantees $(\epsilon,\delta)$, which is the standard criterion in DP, and we additionally report the resulting $\sigma$ in 
Table~\ref{tab:noise_multipliers} for transparency. Although the $\sigma$ values differ across methods, 
these differences reflect their respective mechanisms and composition properties rather than unequal privacy guarantees.}

\begin{table}
    \centering  
    {
    \fontsize{8}{9.5}\selectfont 
    \begin{tabular}{
        >{\centering\arraybackslash}p{1.8cm} |
        >{\centering\arraybackslash}p{1.2cm} |
        >{\centering\arraybackslash}p{1.2cm} |
        >{\centering\arraybackslash}p{1.2cm}}
        \toprule

        \multirow{3}{*}{\textbf{ Method}} &  
        \multicolumn{3}{c}{\textbf{ $\sigma$}} \\
        \cmidrule(lr){2-4}
         & \textbf{$\epsilon$ = 4} & \textbf{$\epsilon$ = 2} & \textbf{$\epsilon$ = 1} \\

        \midrule       
        AUG-PE & 1.07 & 1.97 & 3.68 \\
        \midrule
        DP-SGD & 0.68 & 0.89 & 1.35 \\
        \midrule
        RPSG & 1.20 & 2.40 & 4.80 \\
        \bottomrule
    \end{tabular}
    \caption{\added{Noise Multipliers $\sigma$ at Matched $(\epsilon,\delta)$.}}
    \label{tab:noise_multipliers}
    }   
\end{table}

\begin{figure*}[t]
    \centering
    \begin{minipage}{0.254\textwidth}
        \centering
        \includegraphics[width=\linewidth,height=3.5cm]{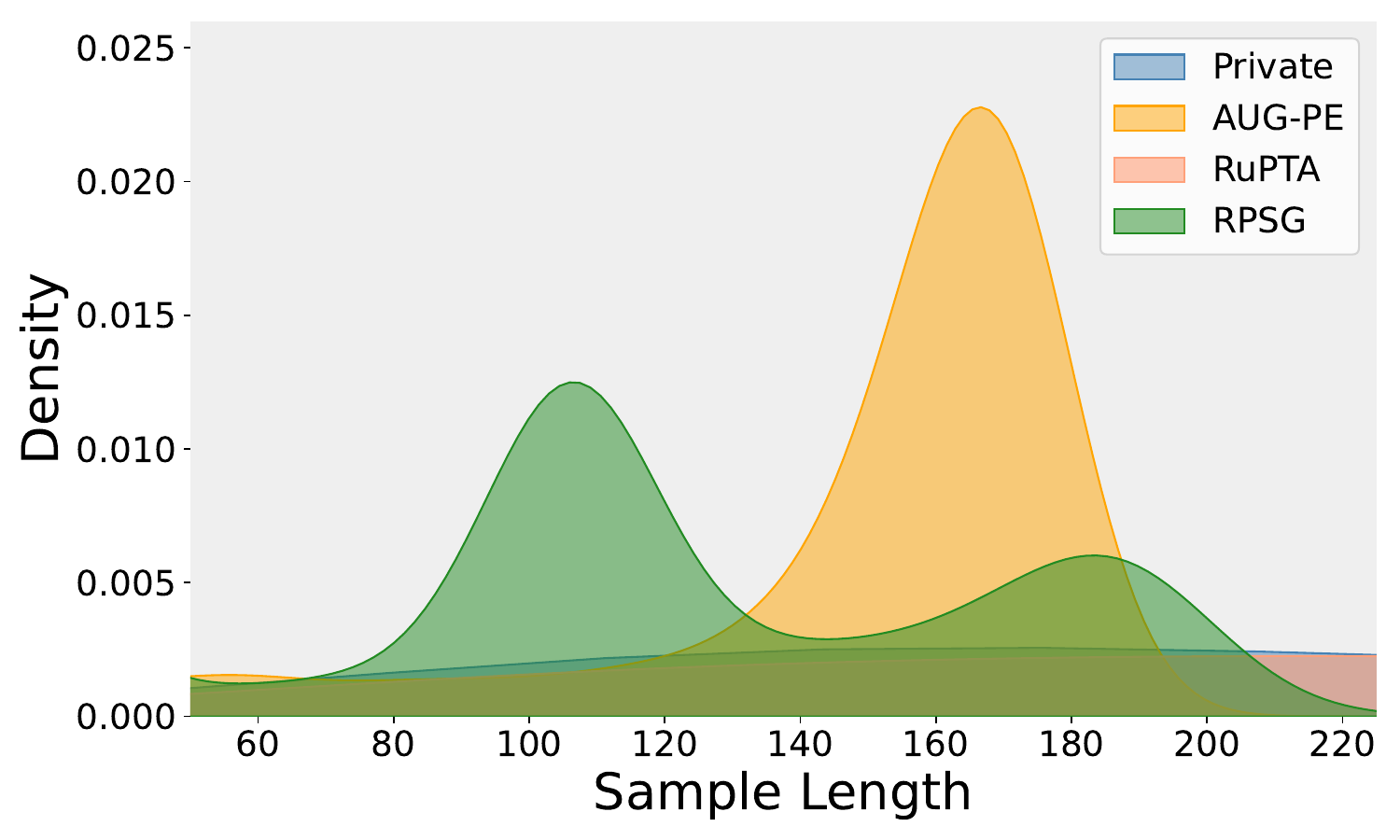}
        \captionsetup{font=footnotesize, labelfont=footnotesize}
        \caption{\added{Length distribution\\of synthetic samples on Reddit with no DP (\(\epsilon=\infty\)).}}
        \label{fig:eval_len}
    \end{minipage}
    \hfill
    \begin{minipage}{0.243\textwidth}
        \centering
        \includegraphics[width=\linewidth,height=3.5cm]{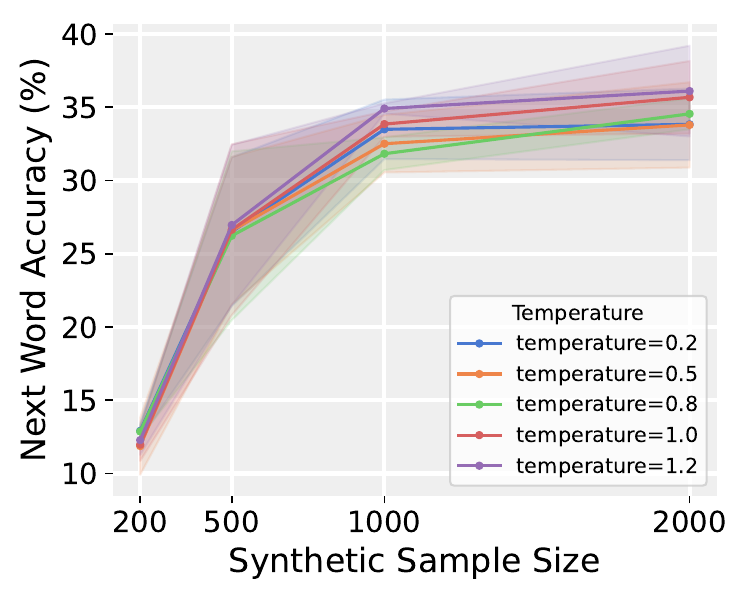}
        \captionsetup{font=footnotesize, labelfont=footnotesize}
        \caption{Effect of synthetic\\sample size and temperature\\on next-word prediction.}
        \label{fig:abla_acc}
    \end{minipage}
    \hfill
    \begin{minipage}{0.243\textwidth}
        \centering
        \includegraphics[width=\linewidth,height=3.5cm]{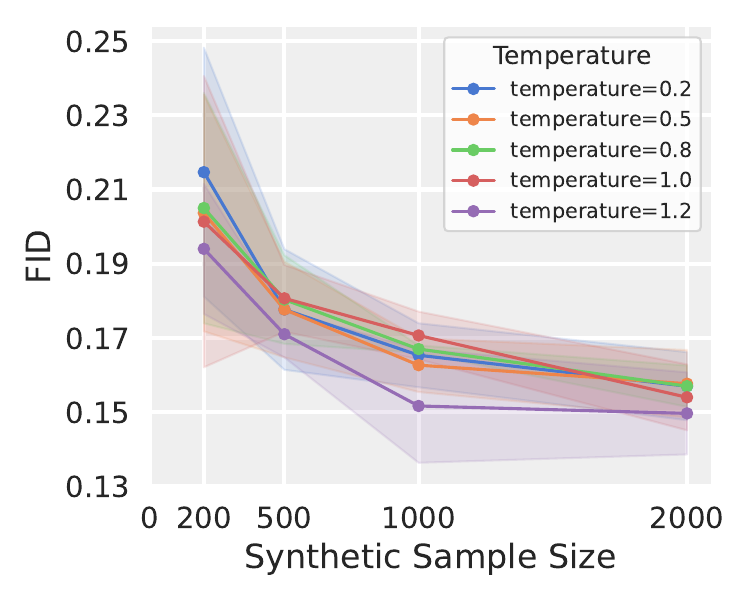}
        \captionsetup{font=footnotesize, labelfont=footnotesize}
        \caption{Effect of synthetic\\sample size and temperature\\on FID.}
        \label{fig:abla_fid}
    \end{minipage}
    \hfill
    \begin{minipage}{0.243\textwidth}
        \centering
        \includegraphics[width=\linewidth,height=3.5cm]{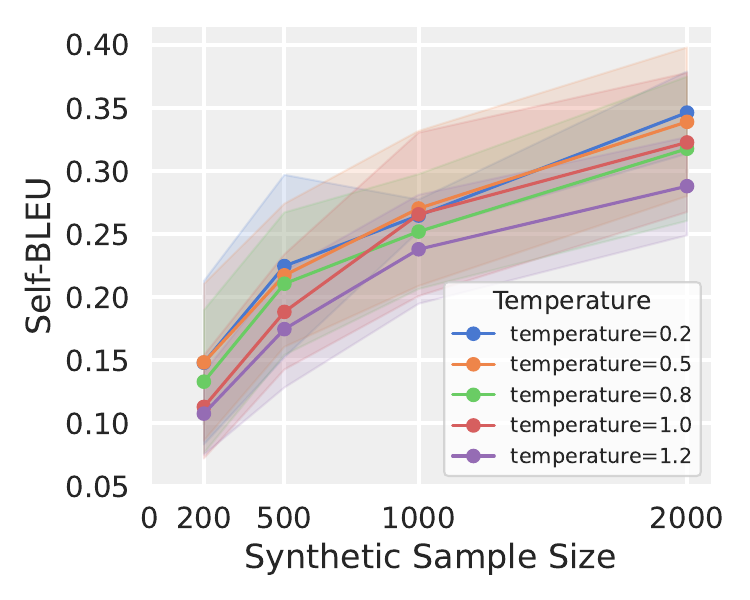}
        \captionsetup{font=footnotesize, labelfont=footnotesize}
        \caption{Effect of synthetic\\sample size and temperature\\on self-BLEU.}
        \label{fig:abla_bleu}
    \end{minipage}
\end{figure*}

\subsection{Synthetic Data Generation Configuration}
\label{appendix-synthetic-data-generation-configuration}
Table~\ref{tab:hyperparameters_filterings} details the experimental configurations used to generate synthetic data across two datasets and six LLMs. The \textit{Initial Synthetic Samples} indicates the number of raw synthetic outputs generated from private seeds before any filtering is applied. These samples undergo two post-generation refinement stages: (1) cosine similarity filtering, which retains a bottom-ranked portion of samples that are less similar to their seed inputs based on embedding cosine similarity. (2) NLL-based filtering, which retains samples with high NLL scores. These two filtering thresholds are expressed as proportions in the \textit{Similarity Threshold} and \textit{NLL Percentile}, indicating the retained portion of samples after each step. The resulting final sample count is reported under \textit{Refined Synthetic Samples}. Other generation hyperparameters, including epoch, learning rate, weight decay, and temperature, are also provided for completeness. The variation in initial synthetic sample sizes across different LLMs reflects both experimental design choices and resource considerations, particularly the cost-related overheads involved in querying API-based models.

\begin{table*}
\centering
\fontsize{9}{10.5}\selectfont
\renewcommand{\arraystretch}{1.5}
\begin{tabular}{>{\centering\arraybackslash}m{1.2cm} | >{\centering\arraybackslash}m{6cm} | >{\centering\arraybackslash}m{6cm}}
\toprule
\textbf{Dataset} & \textbf{Generating Synthetic Variant} & \textbf{Generating Synthetic Data} \\

\midrule
Reddit
& Below is an abstracted self-disclosure statement. Use it to infer the original meaning and rewrite it into a realistic self-disclosure passage:
& Rephrase the following self-disclosure passage into a different but semantically similar version: \\
\midrule
PubMed
& Below is an abstracted abstract of a medical research paper. Rewrite this and preserve the original meaning:
& Rephrase the following sentences as an abstract for medical research paper: \\
\bottomrule
\end{tabular}
\caption{Prompts Used in RPSG to Generate Synthetic Variants and Data.}
\label{tab:prompt_design}
\end{table*}

\begin{table*}
\centering
{
\fontsize{8}{9.5}\selectfont
\begin{tabular}{
    >{\centering\arraybackslash}m{0.8cm}
    >{\centering\arraybackslash}m{1.6cm} |
    >{\centering\arraybackslash}m{1.2cm} |
    >{\centering\arraybackslash}m{0.5cm}
    >{\centering\arraybackslash}m{1.2cm}
    >{\centering\arraybackslash}m{0.9cm}
    >{\centering\arraybackslash}m{0.8cm} |
    >{\centering\arraybackslash}m{1.3cm}
    >{\centering\arraybackslash}m{1.3cm} |
    >{\centering\arraybackslash}m{1.5cm}
}
\toprule
\textbf{Dataset} & 
\textbf{Model} & 
\textbf{Initial Synthetic Samples} & 
\textbf{Epoch} & 
\textbf{Learning Rate} & 
\textbf{Weight Decay} & 
\textbf{Temp} & 
\textbf{Similarity Threshold} & 
\textbf{NLL Percentile} & 
\textbf{Refined Synthetic Samples} \\
\midrule
\multirow{6}{*}{Reddit}
 & GPT-3.5      & 800  & 5 & 3e-4  & 0.01 & 1.0 & 0.4  & 0.4  & 257 \\
 & GPT-4o-mini  & 500  & 5 & 4e-4  & 0.01 & 1.0 & 0.5  & 0.4  & 209 \\
 & DeepSeek-R1  & 350  & 3 & 4e-4  & 0.01 & 1.0 & 0.8  & 0.7  & 195 \\
 & Phi-4        & 2000 & 5 & 4e-4  & 0.01 & 1.0 & 0.65 & 0.55 & 702 \\
 &  DistilGPT2   &  900 &  5  &  5e-4  &  0.01 &  1.0 &  0.85 &  0.3 &  519 \\
 &  GPT-2   &  900 &  5  &  4e-4  &  0.01 &  1.0 &  0.8 &  0.25 &  547 \\
\midrule
\midrule
\multirow{4}{*}{PubMed}
 & GPT-3.5      & 800  & 5 & 1e-4  & 0.01 & 1.0 & 0.45 & 0.35 & 501 \\
 & GPT-4o-mini  & 2500 & 5 & 5e-5  & 0.01 & 1.0 & 0.45 & 0.4  & 1125 \\
 & DeepSeek-R1  & 220  & 3 & 4e-4  & 0.01 & 1.0 & 0.45 & 0.4  & 124 \\
 & Phi-4        & 500  & 5 & 4e-4  & 0.01 & 1.0 & 0.65 & 0.55 & 175 \\
\bottomrule
\end{tabular}
\caption{Hyperparameters and Threshold Settings in our Experiments.}
\label{tab:hyperparameters_filterings}
}
\end{table*}

\begin{table*}
    \centering
    {
    \fontsize{6.5}{6}\selectfont 
    \begin{tabular}{
        >{\centering\arraybackslash}p{0.7cm} 
        >{\centering\arraybackslash}p{1cm} 
        >{\centering\arraybackslash}p{0.9cm} |
        >{\centering\arraybackslash}p{1.4cm}>{\centering\arraybackslash}p{1.2cm}|
        >{\centering\arraybackslash}p{1.4cm}>{\centering\arraybackslash}p{1.2cm}|
        >{\centering\arraybackslash}p{1.4cm}>{\centering\arraybackslash}p{1.2cm}}
        \toprule

        \multirow{3}{*}{\textbf{Dataset}} & 
        \multirow{3}{*}{\textbf{Model}} & 
        \multirow{3}{*}{\textbf{Method}} &  
        \multicolumn{2}{c|}{\textbf{$\epsilon$ = 4}} & 
        \multicolumn{2}{c|}{\textbf{$\epsilon$ = 2}} & 
        \multicolumn{2}{c}{\textbf{$\epsilon$ = 1}} \\
        \cmidrule(lr){4-9}
         & & & 
         \textbf{Self-BLEU($\downarrow$)} & \textbf{N-Gram($\uparrow$)} &
         \textbf{Self-BLEU($\downarrow$)} & \textbf{N-Gram($\uparrow$)} &
         \textbf{Self-BLEU($\downarrow$)} & \textbf{N-Gram($\uparrow$)} \\
        \midrule       
        \multirow{10}{*}{Reddit}  
        & \multirow{2}{*}{GPT-4o-mini} & AUG-PE  
        & 0.65 & 0.28 & 0.64 & 0.29 & 0.63 & 0.29 \\
         &  & RPSG 
        & \textbf{0.41} & \textbf{0.47} & \textbf{0.40} & \textbf{0.47} & \textbf{0.40} & \textbf{0.46} \\
        \cmidrule{2-9}
     
         & \multirow{2}{*}{Phi-4} & AUG-PE  
        & 0.55 & 0.37 & 0.54 & 0.37 & 0.53 & 0.37 \\
         &  & RPSG 
        & \textbf{0.22} & \textbf{0.64} & \textbf{0.24} & \textbf{0.62} & \textbf{0.14} & \textbf{0.76} \\

        \cmidrule[\heavyrulewidth]{2-9}

        & \multirow{2}{*}{DistilGPT2} & DP-SGD  
        & 0.01 & 0.94 & 0.01 & 0.94 & 0.01 & 0.94 \\
         &  & RPSG 
        & 0.19 & 0.57 & 0.21 & 0.55 & 0.20 & 0.55 \\
        \cmidrule{2-9}
     
         & \multirow{2}{*}{GPT-2} & DP-SGD  
        & 0.01 & 0.91 & 0.01 & 0.91 & 0.01 & 0.91 \\
         &  & RPSG 
        & 0.20 & 0.56 & 0.21 & 0.55 & 0.21 & 0.56 \\

        \bottomrule
    \end{tabular}
    \caption{\added{Lexical Diversity Results under Different Privacy Budgets.}}
    \label{tb:lexical_diversity_w_dp}
    }   
\end{table*}

\begin{table*}
    \centering
    { 
    \fontsize{6.5}{6.5}\selectfont 
    \begin{tabular}{>{\centering\arraybackslash}p{0.8cm} >{\centering\arraybackslash}p{1.2cm} >{\centering\arraybackslash}p{1cm} | >{\centering\arraybackslash}p{0.7cm} | >
    {\centering\arraybackslash}p{0.8cm} | >
    {\centering\arraybackslash}p{1cm} | >{\centering\arraybackslash}p{0.8cm} | >{\centering\arraybackslash}p{0.7cm} | >{\centering\arraybackslash}p{0.7cm} | >{\centering\arraybackslash}p{0.7cm} | >{\centering\arraybackslash}p{0.8cm} | >{\centering\arraybackslash}p{0.7cm}}
        \toprule
        \textbf{Dataset} & \textbf{Model} & \textbf{Method} & \textbf{FID}($\downarrow$) & \textbf{Mauve}($\uparrow$) & \textbf{Precision}($\uparrow$) & \textbf{Recall}($\uparrow$) & \textbf{F1}($\uparrow$) & \textbf{KLD}($\downarrow$) & \textbf{TVD}($\downarrow$) & \textbf{WMD}($\downarrow$) & \textbf{SL}($\uparrow$) \\
        \midrule
        \multirow{16}{*}{Reddit} & \multirow{2}{*}{GPT-3.5} 
         & AUG-PE  & 0.15 & 0.01 & 0.95 & 0.15 & 0.25 & 14.2 & 0.99 & 0.01 & 0.10 \\
        &  & RPSG & \textbf{0.09} & \textbf{0.03} & 0.28 & 0.12 & 0.16 & \textbf{3.41} & \textbf{0.83} & 0.02 & \textbf{0.10} \\
        \cmidrule{2-12}
        & \multirow{2}{*}{GPT-4o-mini}
         & AUG-PE  & 0.15 & 0.01 & 0.95 & 0.13 & 0.23 & 13.4 & 0.98 & 0.01 & 0.10 \\
        &  & RPSG & \textbf{0.10} & \textbf{0.02} & 0.84 & \textbf{0.33} & \textbf{0.47} & \textbf{3.93} & \textbf{0.90} & 0.06 & 0.09 \\
        \cmidrule{2-12}
        & \multirow{2}{*}{DeepSeek-R1}
         & AUG-PE  & 0.14 & 0.01 & 0.41 & 0.06 & 0.09 & 7.08 & 0.95 & 0.08 & 0.11 \\
        &  & RPSG & \textbf{0.08} & \textbf{0.12} & 0.39 & \textbf{0.20} & \textbf{0.27} & \textbf{2.99} & \textbf{0.62} & \textbf{0.05} & 0.09 \\
        \cmidrule{2-12}
        & \multirow{2}{*}{Phi-4}
         & AUG-PE  & 0.19 & 0.01 & 0.83 & 0.09 & 0.16 & 13.7 & 0.99 & 0.01 & 0.11 \\
        &  & RPSG & \textbf{0.07} & \textbf{0.25} & 0.53 & \textbf{0.19} & \textbf{0.28} & \textbf{1.41} & \textbf{0.53} & \textbf{0.01} & 0.09 \\

        \cmidrule[\heavyrulewidth]{2-12}
     
         & \multirow{2}{*}{DistilGPT2} &  DP-SGD  
        &  0.10 &  0.22 &  0.04 &  0.05 &  0.05 &  3.40 &  0.57 &  0.01 &  0.13 \\
         &  &  RPSG 
        & \textbf{0.08} &  0.04 & \textbf{0.06} & \textbf{0.09} & \textbf{0.07} & \textbf{2.43} &  0.79 & \textbf{0.01} &  0.12 \\

        \cmidrule{2-12}

         & \multirow{2}{*}{GPT-2} &  DP-SGD 
        &  0.06 &  0.28 &  0.15 &  0.23 &  0.18 &  2.94 &  0.52 &  0.01 &  0.12 \\
         &  &  RPSG 
        & \textbf{0.08} &  0.08 &  0.05 &  0.10 &  0.07 &  5.04 &  0.71 & \textbf{0.01} & \textbf{0.12} \\

        \midrule \midrule
        \multirow{10}{*}{PubMed}  
        & \multirow{2}{*}{GPT-3.5} 
         & AUG-PE  & 0.08 & 0.28 & 0.91 & 0.13 & 0.22 & 0.78 & 0.48 & 0.02 & 0.09 \\
        &  & RPSG & \textbf{0.04} & \textbf{0.85} & 0.82 & 0.08 & 0.15 & \textbf{0.21} & \textbf{0.26} & \textbf{0.01} & \textbf{0.09} \\
        \cmidrule{2-12}
        & \multirow{2}{*}{GPT-4o-mini}
         & AUG-PE  & 0.07 & 0.32 & 0.97 & 0.18 & 0.31 & 0.71 & 0.47 & 0.01 & 0.09 \\
        &  & RPSG & \textbf{0.03} & \textbf{0.72} & 0.71 & 0.07 & 0.13 & \textbf{0.46} & \textbf{0.31} & \textbf{0.01} & \textbf{0.10} \\
        \cmidrule{2-12}
        & \multirow{2}{*}{DeepSeek-R1}
         & AUG-PE  & 0.12 & 0.25 & 0.99 & 0.12 & 0.21 & 0.87 & 0.57 & 0.09 & 0.09 \\
        &  & RPSG & \textbf{0.10} & \textbf{0.87} & 0.94 & 0.05 & 0.09 & \textbf{0.21} & \textbf{0.24} & \textbf{0.04} & \textbf{0.10} \\
        \cmidrule{2-12}
        & \multirow{2}{*}{Phi-4}
         & AUG-PE  & 0.09 & 0.03 & 0.87 & 0.12 & 0.21 & 3.13 & 0.83 & 0.01 & 0.10 \\
        &  & RPSG & \textbf{0.08} & \textbf{0.85} & 0.70 & 0.05 & 0.09 & \textbf{0.24} & \textbf{0.27} & 0.03 & \textbf{0.10} \\
        \bottomrule
    \end{tabular}
    \caption{Distributional and Semantic Similarity to the Private Data for the non-DP Baseline (\(\epsilon=\infty\)).}
    \label{tb:distri_similarity_eval}
    }
\end{table*}

\section{Dataset Construction}
\label{appendix-dataset-construction} 

To build our benchmark Reddit dataset, we scraped posts from Reddit using the \textsc{PRAW} API, focusing on subreddits where users commonly disclose challenges related to poverty and financial instability. The selected subreddits were: \textit{r/frugal}, \textit{r/povertyfinance}, \textit{r/help}, \textit{r/Unemployment}, \textit{r/Assistance}, \textit{r/homeless}, and \textit{r/poverty}. These communities were chosen for their relevance to low socioeconomic populations and their consistent activity.
To ensure the collected posts reflected financial hardship and economic struggle, we applied keyword-based filtering using a curated list of 121 phrases, as shown in Table~\ref{tab:poverty_keywords}.

\begin{table}
    \centering  
    {
    \fontsize{6.5}{6}\selectfont 
    \begin{tabular}{
        >{\centering\arraybackslash}p{0.6cm} 
        >{\centering\arraybackslash}p{0.9cm} 
        >{\centering\arraybackslash}p{0.9cm} |
        >{\centering\arraybackslash}p{0.8cm} |
        >{\centering\arraybackslash}p{0.8cm} |
        >{\centering\arraybackslash}p{0.8cm}}
        \toprule

        \multirow{3}{*}{\textbf{Dataset}} & 
        \multirow{3}{*}{\textbf{Model}} & 
        \multirow{3}{*}{\textbf{Method}} &  
        \textbf{$\epsilon$ = 4} & 
        \textbf{$\epsilon$ = 2} & 
        \textbf{$\epsilon$ = 1} \\
        \cmidrule(lr){4-6}
         &  &  & 
         \textbf{Acc(\%)} &
         \textbf{Acc(\%)} &
         \textbf{Acc(\%)} \\

        \midrule       
        \multirow{10}{*}{Reddit} & \multirow{2}{*}{GPT-4o-mini} & AUG-PE  
        & 32.2 & 31.8 & 32.3 \\
         &  & RPSG 
        & 24.6 & 21.4 & 24.6 \\
        \cmidrule{2-6}
     
         & \multirow{2}{*}{Phi-4} & AUG-PE  
        & 34.6 & 34.7 & 34.5 \\
         &  & RPSG 
        & 29.6 & 30.1 & 32.5 \\

        \cmidrule[\heavyrulewidth]{2-6}
     
         & \multirow{2}{*}{DistilGPT2} & DP-SGD  
        & 4.78 & 4.81 & 5.89 \\
         &  & RPSG 
        & \textbf{34.4} & \textbf{33.9} & \textbf{36.2} \\

        \cmidrule{2-6}

        & \multirow{2}{*}{GPT-2} & DP-SGD  
        & 5.23 & 5.06 & 5.06 \\
         &  & RPSG 
        & \textbf{35.5} & \textbf{36.0} & \textbf{35.5} \\

        \bottomrule
    \end{tabular}
    \caption{ Next-word Prediction Accuracy under Different Privacy Budgets.}
    \label{tb:downstream_eval_w_dp}
    }   
\end{table}

\section{Membership Inference Evaluation: Avoiding Common Pitfalls}
\label{appendix-for-mias}

\subsection{Evaluation Settings}
We carefully designed our experimental pipeline to rigorously evaluate the resistance of our synthetic data to MIAs~\cite{DBLP:conf/uss/CarliniTWJHLRBS21, DBLP:conf/sp/CarliniCN0TT22, DBLP:conf/acl/MatternMJSSB23}. Specifically, our evaluation follows these settings:

\textbf{Dataset Partitioning}: The private dataset(e.g., \textit{train.csv}) is divided into two distinct subsets. The first subset serves as the seed pool for generating synthetic data (members), and the second subset comprises entirely unseen samples (non-members).

\textbf{Synthetic Data Generation}: Synthetic data generation leverages a strict and robust pipeline involving abstraction, sentiment alignment, embedding-based cosine similarity filtering, and NLL screening. This ensures that generated data significantly diverges from direct memorization while preserving utility.

\textbf{Surrogate Model Training}: A freshly initialized BERT-small model (pretrained only) is fine-tuned exclusively on refined synthetic data for five epochs. Given the relatively small synthetic dataset size (typically under 2,000 data points), five epochs provide a realistic opportunity for memorization, rigorously testing privacy robustness.

\textbf{MIAs}: Three dedicated MIAs ~\cite{DBLP:conf/acl/MatternMJSSB23} are employed on the fine-tuned BERT-small model, corresponding to AUC-based metrics: PPL, REFER, and LIRA. Each attack assesses whether a synthetic data sample is more likely to originate from the member or non-member subset of the training data.

\begin{table*}
    \centering  
    {
    \fontsize{6.5}{6}\selectfont 
    \begin{tabular}{
        >{\centering\arraybackslash}p{0.7cm} 
        >{\centering\arraybackslash}p{1cm} 
        >{\centering\arraybackslash}p{0.9cm} |
        >{\centering\arraybackslash}p{0.8cm} >{\centering\arraybackslash}p{0.6cm} >{\centering\arraybackslash}p{0.6cm} |
        >{\centering\arraybackslash}p{0.8cm} >{\centering\arraybackslash}p{0.6cm} >{\centering\arraybackslash}p{0.6cm} |
        >{\centering\arraybackslash}p{0.8cm} >{\centering\arraybackslash}p{0.6cm} >{\centering\arraybackslash}p{0.6cm}}
        \toprule

        \multirow{3}{*}{\textbf{Dataset}} & 
        \multirow{3}{*}{\textbf{Model}} & 
        \multirow{3}{*}{\textbf{Method}} &  
        \multicolumn{3}{c|}{\textbf{AUC ($\epsilon$ = 4)}} & 
        \multicolumn{3}{c|}{\textbf{AUC ($\epsilon$ = 2)}} & 
        \multicolumn{3}{c}{\textbf{AUC ($\epsilon$ = 1)}} \\
        \cmidrule(lr){4-12}
         &  &  & 
         \textbf{PPL} & \textbf{REFER} & \textbf{LIRA} &
         \textbf{PPL} & \textbf{REFER} & \textbf{LIRA} &
         \textbf{PPL} & \textbf{REFER} & \textbf{LIRA} \\

        \midrule       
        \multirow{10}{*}{Reddit} & \multirow{2}{*}{GPT-4o-mini} & AUG-PE  
        & 39.1 & 79.3 & 40.5 
        & 34.5 & 82.9 & 39.5 
        & 57.0 & 61.3 & 44.2 \\
         &  & RPSG 
        & \textbf{53.7} & \textbf{66.5} & 40.1 
        & \textbf{53.1} & \textbf{67.3} & \textbf{40.7} 
        & 64.6 & \textbf{53.7} & \textbf{48.2} \\
        \cmidrule{2-12}
     
         & \multirow{2}{*}{Phi-4} & AUG-PE  
        & 42.1 & 67.4 & 38.1 
        & 46.1 & 62.8 & 41.3 
        & 48.2 & 60.3 & 43.2 \\
         &  & RPSG 
        & \textbf{45.7} & \textbf{62.1} & \textbf{42.3} 
        & \textbf{50.8} & \textbf{55.4} & \textbf{47.3} 
        & 60.2 & \textbf{43.9} & \textbf{53.8} \\

        \cmidrule[\heavyrulewidth]{2-12}
     
         & \multirow{2}{*}{DistilGPT2} & DP-SGD  
        & 52.5 & 39.9 & 57.9 & 50.8 & 34.2 & 60.2 & 52.7 & 44.7 & 54.2 \\
         &  & RPSG 
        & 55.9 & \textbf{54.4} & \textbf{47.0} & 52.9 & \textbf{59.0} & \textbf{43.9} & 60.2 & \textbf{51.0} & \textbf{49.0} \\

        \cmidrule{2-12}

        & \multirow{2}{*}{GPT-2} & DP-SGD  
        & 50.7 & 63.8 & 40.1 & 49.2 & 43.1 & 55.1 & 50.7 & 46.1 & 51.9 \\
         &  & RPSG 
        & 44.8 & 69.5 & 36.5 & 58.8 & \textbf{53.6} & \textbf{47.1} & \textbf{49.5} & 63.4 & 41.5 \\

        \bottomrule
    \end{tabular}
    \caption{\added{Evaluation of MIAs under Different Privacy Budgets.}}
    \label{tb:mia_eval_w_dp}
    }   
\end{table*}

\subsection{Common Methodological Pitfalls Identified in Prior Work}
Recent work by \citet{DBLP:journals/corr/abs-2402-07841} critically evaluates prior MIA methodologies and identifies several key methodological pitfalls that lead to misleading conclusions about the effectiveness of MIAs:

\begin{enumerate}
\item \textbf{Temporal Distribution Shift}: Some studies choose non-members from the same domain (e.g., Wikipedia) but from different temporal snapshots, resulting in artificial temporal shifts rather than true membership signals.
\item \textbf{Artificial Lexical Filtering}: Certain evaluations artificially eliminate overlapping n-grams or lexical similarities between member and non-member samples, creating unnaturally distinguishable datasets.
\item \textbf{Single-Epoch Training}: Evaluating MIAs on models trained for a single epoch on massive datasets inherently limits memorization opportunities, misleadingly suggesting MIAs are ineffective.
\item \textbf{Synthetic Non-member Generation}: Some evaluations generate non-members by applying minimal modifications (e.g., synonyms or paraphrasing) to member samples using LLMs, resulting in semantic overlap and invalidating true membership evaluation.
\end{enumerate}

\subsection{How Our Methodology Avoids These Pitfalls}
Our rigorous approach explicitly avoids each of the pitfall identified above, ensuring the validity and robustness of our MIA evaluations:

\begin{enumerate}
\item \textbf{Avoiding Temporal Distribution Shifts}: Our member and non-member samples are drawn from disjoint subsets of the same data sources, ensuring they represent distinct data points without temporal and content overlap. Since we do not rely on data collected across time windows, our evaluation avoids the confounding effects of temporal drift that can lead to inflated MIA results.

\item \textbf{No Artificial Lexical Filtering}: We do not filter or manipulate member or non-member datasets to reduce lexical overlap. All data samples remain unmodified, preserving natural content appearance and realistic evaluation conditions.

\item \textbf{Realistic Multi-Epoch Training}: Our surrogate BERT-small model is trained for multiple epochs (typically three to five) on relatively small-scale data. This realistic scenario facilitates genuine memorization opportunities, thus providing a stringent test for MIA robustness.

\item \textbf{Valid Non-member Definition}: We define non-members as private samples that were never used to train the surrogate model (i.e., not seen and accessed by BERT-small). Unlike approaches that generate non-members by rephrasing members, our evaluation uses disjoint subsets of private data to ensure a clean membership distinction.
\end{enumerate}

By consciously and carefully avoiding these methodological errors, our results provide a more accurate and meaningful measure of privacy leakage and effectively demonstrate the genuine resistance of our synthetic data to MIAs.

\section{Qualitative Evaluation}
\label{appendix-qualitative-evaluation}

\begin{table*}
\centering
\fontsize{8.5}{9.5}\selectfont
\renewcommand{\arraystretch}{1.3}
\begin{tabular}{>{\centering\arraybackslash}m{1.5cm} | m{13cm} }
\toprule
\textbf{Type} & \multicolumn{1}{c}{\textbf{Reddit}} \\
\midrule
\makecell{Private\\Sample}  
& My \textbf{sister} (25) Myself (25) ( We're \textbf{twins}, same age ) and her \textbf{daughter} who has \textbf{autism} has been \textbf{homeless} for about 1 month now. We're good people and we're clean people. We were \textbf{one month behind on rent} ( rent was \textbf{\$2,200}, it was \textbf{hard paying} that \textbf{month to month} ) and all of sudden everything fell apart so fast. We live in CA in Los Angeles county and these \textbf{shelters} that were provided for us to call etc has been \textbf{sending us in circles}, not calling us back, or saying they have \textbf{no room for us}. It's \textbf{depressing} because getting back on our feet seems \textbf{impossible} because \textbf{all the money is either going to hotels, gas, or fast food and water}. We don't have family anywhere at all or any that cares for us ( trust me we tried \textbf{asking for help} ) I'm wondering has anybody been through the same thing and \textbf{how do your overcome this} ? It's \textbf{so depressing} and now there we're homeless we see how people are treating us different. ...(omitted) \\
\midrule
\makecell{Synthetic\\ Sample\\ (DP-SGD)}
&  Anyway, now that my kid is getting to college and I desperately need to start saving for her, what is the point of it if we all decide that we don lice each other and are going to go around accusing eachother of cheating, of stealing my money or of having a fight with me? Now, if that sounds like you guys, then good on yourself. I'll give you two exceptions. One, because both me and my mom have \textbf{financial issues}, but that proves that you can't just "treat yourself" to some nice decent treatment. Two, being able to afford a place for my family (my wife and kids) doesn't automatically mean that they'll automatically be willing to accommodate you for the rest of your life. Because, to be perfectly honest, those two things are also reasons why I feel like I \textbf{need a shelter}, even temporarily. My shelter will take me to a state border town a thousand miles away, which will let me get to classes, go to the library, and get food. It pays for itself over a period of installments (probably less). ...(omitted) \\
\midrule
\makecell{Synthetic\\Sample\\(AUG-PE)}
& Every day is a struggle on the streets. I wake up in a spot that offers little \textbf{shelter} from the biting cold. With no roof over my head, my body aches, but there’s no money for a doctor. Just a few coins in my pocket — not enough for a bus pass, let alone a hospital visit. I watch people walk by, their lives full of warmth and comfort, while I scrounge for my next meal. Eating feels like a gamble. A half-eaten sandwich tossed in a trash can is often my best bet. I know it’s not healthy, but hunger gnaws at me mercilessly. I used to care about what went into my body, but priorities shift when survival becomes the only goal. Some days, the uncertainty of finding food weighs heavier than the chill in the air. Trying to find work feels \textbf{hopeless}. I can barely clean myself up. Interviews? They require clothes that fit and a sharp presentation, neither of which I possess. The world moves forward while I’m stuck, watching. Every moment I spend \textbf{searching for basic necessities} reminds me of my reality. The \textbf{cycle of homelessness} wraps tightly around me, and there's no escape in sight. \\
\midrule
\makecell{ Synthetic\\ Sample\\ (RUPTA)}
&  My sibling and I (both in our mid-20s) and our niece, who has special needs, have been \textbf{homeless} for about a month now. We're good people and we're clean people. We were \textbf{one month behind on rent} (rent was around \textbf{\$2,000}, it was \textbf{hard paying that month to month}) and all of sudden everything fell apart so fast. We live in a major city in California, and the \textbf{shelters} that were provided for us to call, etc., have been sending us \textbf{in circles}, not calling us back, or saying they have \textbf{no room} for us. It's \textbf{depressing} because getting back on our feet seems \textbf{impossible} because all the money is either \textbf{going to hotels, gas, or fast food and water}. We don't have family anywhere at all or any that cares for us (trust me, we tried \textbf{asking for help}). I'm wondering if anybody has been through the same thing and \textbf{how do you overcome this}? It's so \textbf{depressing}, and now that we're \textbf{homeless}, we see how people are \textbf{treating us differently}. ...(omitted) \\
\midrule
\makecell{Synthetic\\Sample\\(RPSG)}
& \textbf{Every month}, my \textbf{sister} and I \textbf{struggle to pay} our \textbf{rent} of \textbf{[CURRENCY]}. With \textbf{twins} to care for, it feels \textbf{impossible}. Our \textbf{daughter}, who has \textbf{autism}, needs special attention, and we are constantly overwhelmed. Right now, we find ourselves \textbf{homeless}, just trying to get by, and we’re already \textbf{a month behind on payments}. This situation is \textbf{breaking us}, each day a reminder of our struggles. I feel like we're \textbf{slipping through the cracks}, with no one to \textbf{turn to for help}. The weight of \textbf{our circumstances is suffocating}, leaving us feeling \textbf{lost and desperate}. With every passing day, I wonder how we will \textbf{find the means to provide basic necessities} for our \textbf{children}. It’s a painful \textbf{cycle of uncertainty}, and as I look at my \textbf{kids}, all I can think about is how to keep them safe and fed. The fear of \textbf{not knowing where we will sleep at night} is exhausting. Somehow, we must find a way to break free from this \textbf{cycle of despair}. Yet, \textbf{hope feels so far away}. \\
\bottomrule
\end{tabular}
\caption{Example-based Qualitative Comparison Between Private, DP-SGD, AUG-PE, RUPTA, and RPSG Synthetic Samples under $\epsilon=\infty$.}
\label{tab:qual_output_comparison}
\end{table*}

\begin{table*}
    \centering
    {
    \fontsize{7.5}{8.5}\selectfont
    \renewcommand{\arraystretch}{1.1}
    \begin{tabular}{
        >{\centering\arraybackslash}m{0.8cm}
        >{\centering\arraybackslash}m{1.5cm} |
        >{\centering\arraybackslash}m{0.5cm} 
        >{\centering\arraybackslash}m{0.7cm}
        >{\centering\arraybackslash}m{0.6cm} |
        >{\centering\arraybackslash}m{0.6cm}
        >{\centering\arraybackslash}m{0.7cm}
        >{\centering\arraybackslash}m{0.6cm}}
        \toprule
        \multirow{3}{*}{\textbf{Dataset}} &
        \multirow{3}{*}{\textbf{Model}} &
        \multicolumn{3}{c|}{\textbf{AUC (No NLL)}} &
        \multicolumn{3}{c}{\textbf{AUC (With NLL)}} \\
        \cmidrule(lr){3-5}\cmidrule(lr){6-8}
         & & \textbf{PPL} & \textbf{REFER} & \textbf{LIRA} & \textbf{PPL} & \textbf{REFER} & \textbf{LIRA} \\
        \midrule
        \multirow{3}{*}{Reddit}
          & GPT-4o-mini & 100.0 & 0.0   & 0.0   & 52.1 & 59.3 & 43.7 \\
          & DeepSeek-R1 & 5.77  & 98.80 & 48.04 & 52.1 & 58.2 & 45.6 \\
          & Phi-4       & 100.0 & 0.0   & 0.0   & 47.5 & 52.4 & 49.2 \\
        \midrule
        \multirow{3}{*}{PubMed}
          & GPT-3.5     & 100.0 & 0.0   & 0.0   & 50.0 & 42.7 & 57.7 \\
          & GPT-4o-mini & 100.0 & 0.0   & 0.0   & 42.0 & 51.8 & 52.6 \\
          & DeepSeek-R1 & 0.08  & 99.99 & 99.92 & 53.3 & 44.2 & 55.3 \\
        \bottomrule
    \end{tabular}
    \caption{\added{Resistance to MIAs with and without NLL-based Filtering.}}
    \label{tb:mia_nll_ablation}
    }
\end{table*}

\begin{table}
\centering
\fontsize{7}{7}\selectfont
\renewcommand{\arraystretch}{1.3}
\begin{tabular}{>{\centering\arraybackslash}m{1.2cm}|
                >{\centering\arraybackslash}m{1cm}|
                >{\centering\arraybackslash}m{1cm}|
                >{\centering\arraybackslash}m{1cm}|
                >{\centering\arraybackslash}m{1cm}}
\toprule
\textbf{Attribute} & \textbf{AUG-PE} & \textbf{DP-SGD} & \textbf{AUPTA} & \textbf{RPSG} \\
\midrule
Fluency & High &  Moderate &  High & High \\
\cmidrule{1-5}
Semantic Coherence & High &  Low &  High & High \\
\cmidrule{1-5}
Domain Alignment & Moderate &  Low &  High & High \\
\bottomrule
\end{tabular}
\caption{Attribute-level Comparison.}
\label{tab:qualitative_evaluation}
\end{table}

\subsection{Private--synthetic Pair Comparison}
Table~\ref{tab:qual_output_comparison} presents the closest synthetic matches to the private Reddit sample for AUG-PE and \added{DP-SGD (retrieved using cosine similarity), alongside the directly aligned outputs of RUPTA} and RPSG under $\epsilon=\infty$.

The RPSG output preserves highly specific and meaningful elements such as \textbf{sister}, \textbf{twins}, \textbf{daughter}, \textbf{autism}, \textbf{children}, \textbf{homeless}, \textbf{lost and desperate}, \textbf{cycle of uncertainty}, and financial hardship markers like \textbf{a month behind on payments}, \textbf{rent}, \textbf{turn to for help}, \textbf{not knowing
where we will sleep at night}, and \textbf{basic necessities}. These features demonstrate strong semantic and sentiment alignment with the private sample without replicating it directly. 

\added{The DP-SGD outputs illustrate the opposite problem. At $\epsilon \in \{4,2,1\}$, the generations collapse into largely incoherent word salad, reflecting the severe noise introduced to enforce privacy. At $\epsilon=\infty$, the outputs are somewhat more readable but still noisy continuations. Cosine similarity to the private example remained very low ($\leq$ 0.19), suggesting that direct memorization was rare. However, the texts were still too incoherent to be useful for applications such as studying the lived experiences of disadvantaged populations. These results underscore a central limitation of DP-SGD: while it provides formal privacy guarantees, it does so at the cost of downstream utility. }

The AUG-PE sample, while fluent, lacks grounded details such as family structure or specific financial stress, defaulting instead to generic homelessness tropes. As a result, the synthetic data provides little value for downstream analysis.

\added{The RUPTA sample closely mirrors the private narrative and affect, preserving the same core entities and events: sibling and niece relations, one month behind on rent with a rounded dollar amount, shelters sending us in circles with no room, and the repeated emphasis on feeling depressed and seeing no way out. This yields semantic and sentiment alignment comparable to RPSG. However, privacy is weaker: Table~\ref{tb:rpsg_vs_rupta_phi4} shows RUPTA’s MIAs AUCs far from chance (PPL 78.4, REFER 28.6, LIRA 64.0), while RPSG is near 50\% on all three (54.1, 50.9, 50.0). In short, RUPTA matches alignment but does not offer the same resistance to MIAs under this setting.}

\added{Overall, RPSG achieves a better balance, producing outputs that are both privacy-preserving and meaningfully aligned with the semantic , sentiment, and structural features of private data, making them far more suitable for real-world research tasks.}

\subsection{Attribute-level Comparison}

Table~\ref{tab:qualitative_evaluation} summarizes subjective but empirically informed evaluations across three key attributes.

\noindent\textbf{Fluency.} Text produced by \added{AUG-PE, RUPTA, and RPSG is consistently fluent, reflecting their reliance on API-based LLMs. DP-SGD, however, is only moderate in fluency due to the noise introduced during training.}

\noindent\textbf{Semantic Coherence.} \added{RPSG and RUPTA maintain strong semantic coherence, preserving meaningful structure and topic continuity.} AUG-PE is generally coherent but tends to drift toward generic narratives. \added{DP-SGD performs poorly, often producing incoherent or fragmented outputs.}

\noindent\textbf{Domain Alignment.} \added{RPSG and RUPTA both capture domain-specific features effectively.} AUG-PE exhibits weaker alignment, often missing fine-grained terminology, \added{while DP-SGD struggles to retain domain relevance because of quality degradation.}

\section{Computational Efficiency Results}
\label{appendix-computational-efficiency-results}

This evaluation was conducted by measuring GPU hours consumed to generate 1,000 synthetic samples, each approximately 200 words in length, using API-based models (GPT-3.5 and GPT-4o-mini) and open-source models (Phi-4-mini and Phi-4). The experiments were performed on an \textbf{Nvidia} A100 GPU with 40GB VRAM and a \textbf{Cascade Lake} core processor with 128GB RAM. As shown in Figure~\ref{fig:eval_gpu}, RPSG consistently requires fewer GPU hours compared to AUG-PE across the tested advanced models. For the GPT-3.5 model, both AUG-PE and RPSG consume 0.2 GPU hours. For the GPT-4o-mini model, RPSG reduces the GPU hours from 1.1 to 0.9, achieving a speedup of 1.22x. For the Phi-4-mini and Phi-4 models, RPSG reduces the GPU hours from 5.0 to 3.9 and from 6.5 to 4.7, respectively, with corresponding speedups of 1.28x and 1.38x. Overall, RPSG demonstrates consistent efficiency gains, achieving speedups ranging from 1.22x to 1.38x.

One notable observation is the considerable discrepancy in GPU hours between the GPT and Phi model families. GPT models exhibit significantly faster synthetic data generation. Conversely, the Phi models, with smaller parameter counts, display substantially longer runtimes. This disparity can be attributed to differences in architecture optimizations and inference techniques, with GPT models specifically engineered for high-performance generative tasks~\cite{DBLP:journals/corr/abs-2303-08774}. While Phi models demonstrate high efficiency when generating a large number of short sequences, their runtime increases disproportionately when generating longer text samples (e.g., 500 words), suggesting that output length plays a more critical role in computational cost than sample quantity. This behavior is consistent with their technical report, which highlights that Phi models are primarily optimized for reasoning-intensive tasks~\cite{DBLP:journals/corr/abs-2412-08905} rather than high-throughput text generation workloads.

\begin{table}
    \centering  
    {
    \fontsize{7}{6.5}\selectfont 
    \begin{tabular}{
        >{\centering\arraybackslash}m{0.8cm} 
        >{\centering\arraybackslash}m{1cm} |
        >{\centering\arraybackslash}m{1.2cm} |
        >{\centering\arraybackslash}m{0.8cm} |
        >{\centering\arraybackslash}m{1cm}}
        \toprule
        \textbf{Dataset} & \textbf{Model} & \textbf{Sample Size} & \textbf{Ave Len.} & \textbf{Sentiment Align.(\%)} \\    
        \midrule       
        \multirow{4}{*}{Reddit} & \multirow{4}{*}{Phi-4} & 100  & 76 & 86.1 \\
        \cmidrule{3-5}
         &  &  200 & 70 & 89.6 \\ 
         \cmidrule{3-5}
         &  &  500 & 59 & 86.5 \\ 
        \bottomrule
    \end{tabular}
    \caption{\added{Sentiment Alignment Between Private and Abstracted Text.}}
    \label{tb:abstraction_eval}
    }   
\end{table}

\section{Ablation Results}
\label{appendix-ablation-results}

\subsection{\added{NLL-based Filtering and Resistance to MIAs}}
\label{appendix-w/o-nll-mias}

\added{\citet{DBLP:conf/acl/MatternMJSSB23} shows that overfit training samples in language models often have abnormally low token level NLL compared with semantically similar non members, making NLL a useful signal of possible memorization. To show the effect of NLL-based filtering in our pipeline, we report membership inference AUCs before filtering and after filtering.}

\added{Starting with 2,000 synthetic samples generated by Phi-4 on Reddit, a BERT model trained on the full set was highly vulnerable to MIAs, with PPL 100.0, REFER 0.0, and LIRA 0.0, as shown in Table~\ref{tb:mia_nll_ablation}. After applying NLL-based filtering to remove likely overfit samples, 702 examples were retained. A BERT model trained on these filtered samples showed much stronger privacy, with PPL 47.5, REFER 52.4, and LIRA 49.2. This reduction from 2,000 to 702 is reflected in Table~\ref{tab:hyperparameters_filterings}.}

\added{Across datasets and LLMs, removing likely overfit samples with NLL-based filtering moves AUCs toward 50\%, often from highly vulnerable values near 0 or 100. This confirms that NLL-based filtering is necessary for privacy robustness in our setting.}

\subsection{Effect on Next-word Prediction Accuracy}
\label{appendix-ablation-accuracy}

As shown in Figure~\ref{fig:abla_acc}, synthetic sample size is the primary factor affecting next-word prediction accuracy. At 200 samples, accuracy is relatively low (around 12–13\%), indicating significant underfitting due to insufficient data. Increasing the sample size from 200 to 500 markedly improves accuracy to approximately 30\%, reflecting enhanced exposure to language patterns. Further increases to 1000 and ultimately 2000 samples yield additional improvements, bringing accuracy close to 40\%.

The saturation near 40\% indicates a fundamental trade-off inherent in the RPSG method. While RPSG utilizes private data as seeds, enabling synthetic samples to closely match authentic language patterns and potentially achieve high accuracy, the refinement procedure intentionally filters out highly similar and memorized samples. This step, essential for robust privacy protection, inevitably excludes samples that could enhance next-word prediction performance. Thus, the observed leveling-off of accuracy clearly illustrates the intrinsic privacy-utility balance within RPSG.

Temperature variations (from 0.2 to 1.2) show minimal influence on accuracy across all synthetic sample sizes, likely due to the specific characteristics of Phi-4. Unlike general-purpose models such as the GPT series, Phi-4 is optimized for reasoning-centric tasks, producing stable, logically coherent outputs. Consequently, changing the temperature, which typically modulates generation diversity, minimally affects Phi-4’s downstream next-token prediction performance. This indicates that, for reasoning-focused models like Phi-4, the quantity of synthetic samples plays a more critical role than temperature variations.

Importantly, despite constraints imposed by privacy considerations, the accuracy of RPSG-generated synthetic data consistently surpasses that of the AUG-PE baseline, as detailed further in Table~\ref{tb:downstream_eval}, highlighting RPSG’s effectiveness in maintaining downstream task performance even under stringent privacy considerations.

\subsection{Effect on FID}

Figure~\ref{fig:abla_fid} demonstrates that FID scores consistently improve (decrease) as synthetic sample size increases across all temperature settings. Larger datasets allow synthetic distributions to better align structurally and semantically with private data. Additionally, FID improves slightly as temperature increases from 0.2 to 1.2, suggesting that more diverse generation (higher temperature) helps synthetic samples better represent the private data distribution, especially with limited sample sizes.

However, improvements from increasing sample size are notably greater than those from adjusting temperature. This observation may stem from Phi-4's strong optimization towards generating consistent, structured semantic outputs, making it relatively insensitive to temperature variations regarding distributional alignment. These results highlight the significance of synthetic dataset size in improving fidelity and suggest that higher temperatures can moderately enhance semantic alignment with private data.

\subsection{Effect on Self-BLEU}
Figure~\ref{fig:abla_bleu} shows self-BLEU scores increasing (e.g., from around 0.1 to 0.3) with larger synthetic sample sizes, indicating reduced diversity. This reduction arises from repetitive phrase structures and recurring patterns becoming more common with larger datasets. Conversely, higher temperatures reduce self-BLEU scores, reflecting increased lexical and syntactic diversity.

This observed pattern matches our expectations: lower temperatures generate more coherent but less diverse outputs, while higher temperatures increase diversity at the potential expense of coherence. Phi-4 clearly demonstrates this behavior.

\subsection{Conclusion of Ablation}
Overall, these ablation results emphasize synthetic sample size as the most influential factor for model accuracy and distributional fidelity. Temperature has limited impact on accuracy and moderate effects on diversity (self-BLEU) and distributional alignment (FID). Phi-4 is notably less sensitive to temperature changes compared to general-purpose LLMs like GPT-4o. Complete results are provided in Table~\ref{tb:ablation_result}.

\begin{table*}
    \centering
    {
    \fontsize{7}{7}\selectfont
    \begin{tabular}{
        >{\centering\arraybackslash}p{0.8cm} 
        >{\centering\arraybackslash}p{1.2cm} | 
        >{\centering\arraybackslash}p{0.7cm} | >{\centering\arraybackslash}p{0.7cm} | >{\centering\arraybackslash}p{0.7cm} | 
        >{\centering\arraybackslash}p{0.7cm} | >{\centering\arraybackslash}p{0.7cm} | >{\centering\arraybackslash}p{0.7cm} | 
        >{\centering\arraybackslash}p{0.7cm} | >{\centering\arraybackslash}p{0.7cm} | >{\centering\arraybackslash}p{0.7cm}
        }
        \toprule
        \textbf{Sample} & \textbf{Temp} & 
        \multicolumn{3}{c|}{\textbf{Accuracy(\%)}} & 
        \multicolumn{3}{c|}{\textbf{FID}} & 
        \multicolumn{3}{c}{\textbf{Self-BLEU}} \\
        \cmidrule(lr){3-5} \cmidrule(lr){6-8} \cmidrule(lr){9-11}
        \textbf{Size} &  & \textbf{Run1} & \textbf{Run2} & \textbf{Run3} & \textbf{Run1} & \textbf{Run2} & \textbf{Run3} & \textbf{Run1} & \textbf{Run2} & \textbf{Run3} \\
        \midrule
        \multirow{5}{*}{200}
        & 0.2 & 13.16 & 13.28 & 12.28 & 0.182 & 0.213 & 0.249 & 0.222 & 0.12 & 0.102 \\
        & 0.5 & 12.36 & 9.72 & 13.53 & 0.175 & 0.198 & 0.238 & 0.22 & 0.114 & 0.111 \\
        & 0.8 & 12.98 & 12.43 & 13.18 & 0.175 & 0.203 & 0.237 & 0.195 & 0.12 & 0.084 \\
        & 1.0 & 12.28 & 10.77 & 12.16 & 0.169 & 0.19 & 0.245 & 0.159 & 0.099 & 0.081 \\
        & 1.2 & 11.2 & 13.21 & 12.42 & 0.176 & 0.195 & 0.211 & 0.144 & 0.098 & 0.081 \\
        \midrule
        \midrule
        \multirow{5}{*}{500}
        & 0.2 & 29.89 & 28.95 & 20.66 & 0.165 & 0.172 & 0.196 & 0.308 & 0.187 & 0.179 \\
        & 0.5 & 29.44 & 29.52 & 20.76 & 0.167 & 0.174 & 0.192 & 0.282 & 0.194 & 0.176 \\
        & 0.8 & 29.44 & 29.66 & 19.57 & 0.172 & 0.175 & 0.194 & 0.273 & 0.196 & 0.163 \\
        & 1.0 & 29.4 & 30.57 & 19.93 & 0.18 & 0.172 & 0.19 & 0.241 & 0.164 & 0.16 \\
        & 1.2 & 29.91 & 30.32 & 20.63 & 0.169 & 0.166 & 0.178 & 0.226 & 0.162 & 0.136 \\
        \midrule
        \midrule
        \multirow{5}{*}{1000}
        & 0.2 & 34.79 & 31.15 & 34.53 & 0.156 & 0.167 & 0.173 & 0.276 & 0.267 & 0.251 \\
        & 0.5 & 34.02 & 33.21 & 30.29 & 0.159 & 0.158 & 0.171 & 0.341 & 0.24 & 0.23 \\
        & 0.8 & 33.04 & 30.85 & 31.58 & 0.168 & 0.166 & 0.167 & 0.302 & 0.241 & 0.213 \\
        & 1.0 & 34.05 & 32.9 & 34.59 & 0.168 & 0.166 & 0.178 & 0.283 & 0.194 & 0.32 \\
        & 1.2 & 34.98 & 34.53 & 35.22 & 0.161 & 0.16 & 0.134 & 0.247 & 0.191 & 0.276 \\
        \midrule
        \midrule
        \multirow{5}{*}{2000}
        & 0.2 & 35.46 & 31.02 & 35.01 & 0.147 & 0.165 & 0.159 & 0.384 & 0.328 & 0.327 \\
        & 0.5 & 35.36 & 30.43 & 35.58 & 0.148 & 0.166 & 0.159 & 0.407 & 0.304 & 0.306 \\
        & 0.8 & 34.89 & 35.39 & 33.26 & 0.152 & 0.163 & 0.156 & 0.383 & 0.293 & 0.277 \\
        & 1.0 & 35.3 & 33.41 & 38.32 & 0.151 & 0.164 & 0.147 & 0.356 & 0.259 & 0.353 \\
        & 1.2 & 35.48 & 33.36 & 39.46 & 0.151 & 0.16 & 0.138 & 0.323 & 0.246 & 0.295 \\
        \bottomrule
    \end{tabular}
    \caption{Ablation results on Next-Word Prediction Accuracy, FID, and self-BLEU across temperature and synthetic sample size settings using Phi-4.}
    \label{tb:ablation_result}
    }
\end{table*}

\begin{table*}
    \centering
    {
    \fontsize{7.5}{8.5}\selectfont
    \renewcommand{\arraystretch}{1.2}
    \begin{tabular}{>{\raggedright\arraybackslash}p{4cm} >{\raggedright\arraybackslash}p{4cm} >{\raggedright\arraybackslash}p{4cm}}
        \toprule
        low income & financial burden & poverty \\
        poor & unemployment & underemployment \\
        wage stagnation & low wage & minimum wage \\
        underpaid & low pay & barely making ends meet \\
        paycheck to paycheck & financial hardship & money struggles \\
        heavy debt & burdened by debt & bankrupt \\
        financially strained & affordability issues & can’t afford \\
        cannot afford & not afford & unable to afford \\
        can’t even afford & cannot even afford & not even afford \\
        unable to even afford & can’t make rent & cannot make rent \\
        not make rent & unable to make rent & can’t pay \\
        cannot pay & not pay & unable to pay \\
        can’t even pay & cannot even pay & not even pay \\
        unable to even pay & broke & financial stress \\
        falling behind on bills & foreclosure & eviction \\
        homeless & food insecurity & social assistance \\
        food stamps & food bank & unable to afford healthcare \\
        living in poverty & working poor & underprivileged \\
        financially disadvantaged & impoverished & economic inequality \\
        barely scraping by & drowning in debt & completely broke \\
        struggling to make ends meet & unable to pay rent & living with no savings \\
        behind on mortgage payments & at risk of eviction & losing access to healthcare \\
        living without health insurance & relying on public assistance & dependent on welfare programs \\
        surviving on food stamps & visiting food banks & facing job insecurity \\
        unable to find stable work & working multiple low-wage jobs & overwhelmed by medical bills \\
        drowning in credit card debt & falling behind on utility payments & struggling with student loan debt \\
        on the verge of bankruptcy & forced to skip meals & cutting back on basic necessities \\
        relying on payday loans & trapped in a cycle of poverty & receiving unemployment benefits \\
        depending on disability checks & living in low-income housing & experiencing financial instability \\
        living in a shelter & surviving on minimum wage & forced to sell personal belongings \\
        dealing with repossession of property & living below the poverty line & burdened by medical debt \\
        struggling to keep utilities on & receiving government assistance & struggling with mental health \\
        facing discrimination in housing & need the money & stuck in a dead-end job \\
        barely surviving & skipping meals to save money & living off handouts \\
        behind on bills & living off ramen & maxed out credit cards \\
        in over my head with debt & living off food stamps & crippling student loans \\
        living in my car & working two jobs just to survive & no health insurance \\
        racked up medical debt & shutoff notice for utilities & about to be homeless \\
        selling my stuff to get by & living off unemployment & struggling to pay rent \\
        working minimum wage & eviction notice came & utilities about to be shut off \\
        credit score tanked \\
        \bottomrule
    \end{tabular}
    \caption{Phrases Used to Filter Posts During Construction of the Reddit Dataset.}
    \label{tab:poverty_keywords}
    }
\end{table*}
\end{document}